\documentclass[letterpaper,twocolumn,10pt]{article}
\usepackage{usenix2019_v3}
\usepackage[english]{babel}
\usepackage[utf8]{inputenc}
\usepackage{url}
\usepackage{caption}
\usepackage{subcaption}
\usepackage{hhline}
\usepackage{pgf}
\usepackage{changepage}
\usepackage{listings}
\usepackage{appendix}
\usepackage{diagbox}
\lstdefinestyle{mystyle}{
    basicstyle=\footnotesize,
    showtabs=false,                  
    tabsize=2
}
\usepackage{enumitem}
\usepackage{graphicx}
\usepackage{xcolor,colortbl}
\usepackage{tikz}
\usepackage{longtable}
\usepackage{mathtools}

\usepackage{abraces}
\usepackage{tabu, multirow, wrapfig, boldline, makecell, booktabs, amsmath}
\usetikzlibrary{decorations.pathreplacing}
\usepackage[ruled,vlined]{algorithm2e}
\usepackage{float}
\usepackage{bytefield}
\usetikzlibrary{positioning}
\usetikzlibrary{shapes,snakes,arrows}
\usepackage{verbatim}
\tikzstyle{process} = [rectangle, minimum width=3cm, minimum height=1cm, text centered, text width=3cm, draw=black, fill=orange!30]
\tikzstyle{cloud} = [draw, ellipse,fill=blue!20, node distance=3cm,
    minimum height=1em]
\tikzstyle{rect} = [draw, rectangle, fill=blue!20, node distance=2cm,
   text centered, text width=4em]
\tikzstyle{block} = [rectangle, draw, fill=blue!20, 
    text centered, rounded corners, minimum height=1em]
\tikzstyle{decision} = [diamond, text centered, draw=black, fill=blue!20,scale=0.5] 
\tikzstyle{line} = [draw, ->]
\tikzset{main node/.style={circle,draw,inner sep=0pt}} 
\usepackage{authblk}
\newboolean{Longversion}
\setboolean{Longversion}{false}

\title{\Large \bf LB Scalability: Achieving the Right Balance Between Being Stateful and Stateless}
\date{}

\begin{document}
\author[1]{Reuven Cohen}
\affil[1]{Department of Computer Science, Technion, Israel}

\author[2]{Matty Kadosh}
\affil[2]{Mellanox, Israel}

\author[2]{Alan Lo}

\author[1]{Qasem Sayah}

\maketitle
\begin{abstract}
A high performance Layer-4 load balancer (LB) is one of the most important components of a cloud service infrastructure. Such an LB uses network and transport layer information for deciding how to distribute client requests across a group of servers. A crucial requirement for a stateful LB is per connection consistency (PCC); namely, that all the packets of the same connection will be forwarded to the same server, as long as the server is alive, even if the pool of servers or the assignment function changes. The challenge is in designing a high throughput, low latency solution that is also scalable.
\par This paper proposes a highly scalable LB, called Prism, implemented using a programmable switch ASIC. As far as we know, Prism is the first reported LB that can process millions of connections per second and hundreds of millions connections in total, while ensuring PCC. This is due to the fact that Prism forwards all the packets in hardware, even during server pool changes, {\bf while avoiding the need to maintain a hardware state per every active connection}. We implemented a prototype of the proposed architecture and showed that Prism can scale to 100 million simultaneous connections, and can accommodate more than one  pool update per second.
\end{abstract}
\vspace{-0.1cm}
\section{Introduction}
\label{sec:introduction}
Network load balancing is a technology area that garners great deal of attention due to the pervasiveness of cloud datacenters. A high performance load balancer (LB) is a crucial component of cloud service infrastructure \cite{patel2013ananta,vaquero2011dynamically}. Layer-4 load balancing uses both network and transport layer information for deciding how to distribute client requests across a group of servers.

\par A cloud service can be identified by a virtual IP (VIP) address. Each VIP is mapped to a pool of servers, and each server is uniquely identified  using a direct IP (DIP) address. The load balancing logic distributes client connections destined for a certain VIP to one of the DIPs associated with this VIP. The distribution is usually performed using a uniform hash function. For TCP connections, this assignment function is calculated using the packet headers 5-tuple, also known as the "connection key". The key is identical for all the packets of the same connection, and therefore all these packets are delivered to the same DIP.

\par One of the most important requirements for an LB is per-connection consistency (PCC) \cite{govindan2016evolve}. PCC means that all packets of the same TCP connection are forwarded to the same DIP as long as this DIP does not malfunction, even in the event of a DIP pool change. Such a change happens when more DIPs are added to the pool, when some DIPs have to be gracefully taken down, or when the assignment function has to be changed.

\par Layer-4 LBs are built for highly scalable environments where millions of connections arrive every second, and DIP pool updates happen frequently \cite{govindan2016evolve}. Violating PCC results in broken connections. Re-establishing a broken connection by the application often takes a long time, and may result in a loss of revenue to the cloud operator \cite{firestone2018azure}. Furthermore, broken connections can violate the service level agreement between the cloud operator and its customers \cite{patel2009service, vaquero2008break}.

\par It is not easy to ensure PCC even if the LB is stateful, namely, it maintains a state per connection. States are maintained by mapping a connection key
(the 5-tuple of the TCP/IP header, or a digest thereof) to the DIP
selected for that connection. An example of a stateful LB is SilkRoad~\cite{miao2017silkroad}, which stores all connection states in a hardware ConnTable to preserve forwarding performance. Inserting a new connection state into the ConnTable is relatively slow compared to the packet arrival rate. Thus, a connection state might not be ready when the pool is updated. This is a critical issue since pool updates occur frequently; e.g., 50 updates per minute in some clusters of large web service provider \cite{miao2017silkroad}. SilkRoad addresses this potential problem by using a special hardware table, which implements a Bloom filter, and maintains the signature of pending connections, namely, those whose SYN has been received, but their state has not yet entered into ConnTable.

\par In contrast to stateful LBs, stateless LBs do not maintain a state per active connection. Thus, when such LBs are used, it is even harder to guarantee PCC. One solution, presented in \cite{olteanu2018stateless}, proposes that the state will be maintained by the DIPs, and that the DIPs will redirect to each other received packets that do not belong to them. For example, if a newly added server receives a packet of an existing connection following a server pool update, it redirects the packet to the correct DIP. This requires strong coordination between the DIPs and adds latency and overhead to the forwarding of each affected packet.

\par Ensuring PCC is easier for software LBs \cite{eisenbud2016maglev,patel2013ananta} because every packet is processed by the software, and race conditions between hardware and software are more easily prevented. However, software LBs have several major drawbacks compared to their hardware counterparts: low throughput, high latency, and high cost \cite{gandhi2015duet,eisenbud2016maglev,guo2015pingmesh}.

\par A scarce resource in a switch-based hardware LB is memory, which is limited to hundreds MB of SRAM in modern switching ASICs \cite{spectrum,tofino,tomahawk}. Storing the state for ten million concurrent connections takes a few hundreds MB of SRAM. For this reason, keeping a hardware state per connection significantly limits the scalability of a hardware LB.

\par Two approaches have been recently proposed for hardware LBs that ensure PCC. The first is SilkRoad \cite{miao2017silkroad}, which maintains a hardware state per active connection. As explained above, the challenge with this approach is that it strictly limits the number of connections concurrently served by the LB. The second approach is Duet \cite{gandhi2015duet}, which maintains a software connection table. The problem with this approach is that in order to update the DIP pool of a VIP while ensuring PCC, Duet needs to redirect all the traffic of that VIP to software processing. This increases the latency of the affected connections, and requires significant CPU resources \cite{miao2017silkroad}.

\par Our LB, Prism, takes a big step forward compared to SilkRoad and Duet, while benefiting from the advantages of both.
Like SilkRoad, Prism processes all packets in hardware, before and after a pool change. But unlike SilkRoad, Prism does not need a hardware state per connection. Like Duet, Prism does not keep a hardware state per active connection. But unlike Duet, Prism processes all packets -- before and after a pool update -- in hardware. \ifLongversion The above discussion is summarized by Table~\ref{table:compare}.

\begin{table}[t]
\begin{center}
\footnotesize
\newcolumntype{C}[1]{>{\centering\arraybackslash}m{#1}}
\newcolumntype{?}{!{\vrule width 1pt}}
\begin{tabular}{ | c | C{2.5cm}| C{3cm} | } 
  \hline
  LB &  Scale & processing packets of affected connections  \\
  \hhline{|=|=|=|}
  \multirow{3}{1.6cm}{\centering Prism (this\;work)}  &  \multirow{3}{2.5cm}{\centering a hardware state only for affected connections}   &  \multirow{3}{*}{by hardware}  \\
   &  & \\
   &  & \\
  \hline
  SilkRoad  & \multirow{2}{2.6cm}{\cenering a hardware state for all connections} & \multirow{2}{*}{by hardware} \\
  \cite{miao2017silkroad} &  &\\
  \hline
  Duet  & \multirow{2}{*}{none} & \multirow{2}{*}{by software} \\ 
  \cite{gandhi2015duet} & &\\
  \hline
\end{tabular}
\caption{ LB comparison}
\label{table:compare}
\end{center}
\vspace{-1cm}
\end{table}
\fi 
\par Since Prism does not maintain a hardware state per connection, it can handle bursts of 1 million connections per second, and a total number of active connections in the hundreds of millions. {\bf As far as we know, no previous reported LB has similar performance and scale. }

\par The rest of this paper is organized as follows. Section~\ref{high-level description} and Section~\ref{sec:detailed} present a high-level and low-level description of Prism's architecture, focusing on the various hardware tables. Section~\ref{sec:dip pool update} discusses DIP pool update and presents Prism's PCC algorithm. Section~\ref{sec:signatures} discusses connection signature, collision issues, the detection and removal of dead connections, and SYN flood attacks. Section~\ref{sec:prototype} provides our Proof of Concept (POC) implementation. Section~\ref{sec:related_work} discusses related work, and Section~\ref{sec:conclusion} concludes the paper. Appendix~\ref{sec:choosing bins} discusses optimizations related to Prism's PCC algorithm. Appendix~\ref{app:p4_implementation} presents our P4 implementation for the various tables and their actions.

 \ifLongversion
 \par Table~\ref{tab:acronyms_table} presents the acronyms used throughout the paper.
 \fi
 
\section{High-Level Description and General Packet Flow}
\label{high-level description}
\ifLongversion
SilkRoad \cite{miao2017silkroad} is a recently proposed layer-4 LB which stores in the hardware a per-connection state. The state is maintained in a table called ConnTable, which maps a connection key (the 5-tuple of the TCP/IP header, or a part thereof) to the DIP selected for that connection. 

Another hardware table, called VIPTable, maps between each VIP to its DIP pool. Each incoming packet is first checked against ConnTable. If there is a hit, the packet is forwarded to the DIP indicated by the ConnTable entry. If there is no hit, as is the case for the first TCP (SYN) packet of a connection, and for all following packets that are received by the LB before the connection is inserted into the ConnTable, the packet is forwarded to the VIPTable, which indicates the chosen DIP for this connection. The mapping between a connection and its chosen DIP is then inserted by the software into the ConnTable. Afterwards, packets of the this connection find a hit in the ConnTable, and are forwarded according to the mapping determined by the VIPTable for the first packet, even if this mapping is changed later.\fi

\ifLongversion
\begin{table}[t]
\begin{center}
\begin{tabular}{ | c | c | } 
  \hline
  Acronym & Meaning \\
  \hlineB{2.5}
  LB & Load Balancer  \\
  \hline
  VIP & Virtual IP Address  \\
  \hline
  DIP & Direct IP Address  \\ 
  \hline
  PCC & Per-Connection Consistency   \\
  \hline
    MCT & Migrated Connection Table   \\
  \hline
    SCT & Software Connection Table   \\
  \hline
    ECMP & Equal-Cost Multi-Path   \\
  \hline
\end{tabular}
\caption{ Abbreviations and acronyms}
\label{tab:acronyms_table}
\end{center}
\vspace{-1cm}
\end{table}
\fi

\par 
 In hardware LBs such as SilkRoad \cite{miao2017silkroad},  PCC can be violated when the DIP pool is updated, because inserting into ConnTable a new mapping between a connection and its DIP is not an atomic operation. For example, suppose that a TCP SYN packet of a new connection  is received and the destination VIP is mapped to a certain DIP. Suppose also that the DIP pool corresponding to the destination VIP is being updated before the new connection is added into the ConnTable. Consequently, the DIP that receives the first packets might not be the one that receives the next packets of this connection.

\par SilkRoad ensures PCC using another special hardware table, called TransitTable. This table remembers the keys of the connections that might violate PCC; i.e., those that should be mapped according to the old pool, but their insertion into the ConnTable is not completed before the pool is updated. \ifLongversion When a packet misses on ConnTable and its VIP pool is being updated, the packet will find a match on the TransitTable, and will be forwarded according to the old server pool and not the new one. The TransitTable is implemented as a Bloom filter, due to its memory efficiency and fast hardware insertion. The use of a Bloom filter introduces false positives. To address such exceptions SilkRoad forwards to the software any TCP SYN packet that matches on the TransitTable, but the number of these packets is expected to be negligible.\fi

\par A major challenge for SilkRoad is that it requires a hardware state in ConnTable for each active connection. This design limits the number of connections that can be simultaneously maintained. In contrast to SilkRoad, Prism does not keep a hardware state for each connection, but only for those whose mapping changes following a pool update. The expected number of such connections is significantly smaller compared to the total number of active connections. E.g. if the system DIP update rate is {\bf 1 every second}, Prism needs a hardware state only for {\bf 0.03\%} of the connections (see Section \ref{sec:dip pool update}). In this manner, Prism can support hundreds of million active connections with arrival rate of 1 million per second, while SilkRoad can support a total of 10M connections \cite{miao2017silkroad}. Additionally, Prism uses only a small fraction (10\%) of the switch's SRAM, leaving enough resources for other important switch functionality, such as routing and tunneling. 

\begin{figure}[t]
\centering
\begin{tikzpicture}[node distance = 2cm, auto] 

\node (dec) [decision, text width=0.6cm] {\small SYN FIN RST};

\node [cloud, above= 0.5cm of dec] (packet) {\tiny packet};

\node (mct) [right = 0.5 cm of dec]{
\scalebox{0.6}{
\begin{tabular}{ | c | c | }
  \hline
  \multicolumn{2}{|c|}{MCT} \\
  \hline
  connection signature & DIP \\
  \hhline{|=|=|}
  conn1-signature & DIP1  \\
  \hline
  conn2-signature & DIP2  \\ 
  \hline
\end{tabular}}};

\node[main node, right = 0.7cm of mct] (hash) {\tiny $Hash$};

\node (ecmp)  [right = 0.3cm of hash]{
\scalebox{0.6}{
\begin{tabular}{ | c | } 
  \hline
  ECMP Table \\
  of VIP1 \\
  \hhline{|=|}
  1. DIP1  \\
  \hline
  2. DIP2  \\
  \hline
  3. DIP3  \\ 
  \hline
  4. DIP4  \\ 
  \hline
  $\vdots$ \\
  \hline
  125. DIP1  \\
  \hline
  126. DIP2  \\
  \hline
  127. DIP3  \\ 
  \hline
  128. DIP4  \\ 
  \hline
\end{tabular}}};

\node (fib) [below = 0.4cm of mct] {
\scalebox{0.6}{
\begin{tabular}{ | c | c | }
  \hline
  \multicolumn{2}{|c|}{FIB(SYN/FIN/RST)} \\
  \hline
  VIP & connection signature \\
  \hhline{|=|=|}
  VIP3 & conn3-signature  \\ 
  \hline
  VIP4 & conn4-signature   \\ 
  \hline
\end{tabular}}
};
\node [block, above of= mct] (forward) {\tiny forward packet};
\path [line] (packet) -- (dec);
\node (ee) [xshift=0.3cm] at (mct.west){};
\path [line] (dec) -- node {\tiny no} (ee);
\node (dd) [xshift=0.3cm] at (fib.west){};
\path [line] (dec.south) |- node[xshift=0.8cm] {\tiny yes} (dd);
\node (cc) [xshift=-0.25cm] at (mct.east){};
\path [line] (cc) -- node {\tiny miss} (hash);
\node (aa) [yshift=0.25cm] at (mct.south){};
\node (bb) [yshift=0.42cm] at (fib.center){};
\path [line] (bb) -- (aa);
\node (ff) [yshift=-0.25cm] at (mct.north){};
\path [line] (ff) -- node {\tiny hit} (forward);
\node (gg) [yshift=-0.25cm] at (ecmp.north){};
\path [line] (gg) |- (forward);
\path [line] (hash) -- (5.65,0.6);
\path [line] (hash) -- (5.65,0);
\path [line] (hash) -- (5.65,-0.6);
\end{tikzpicture}
\caption{\centering Packet flow for a specific VIP (VIP1)}
\label{packet_flow_figure}
\vspace{-0.5cm}
\end{figure}
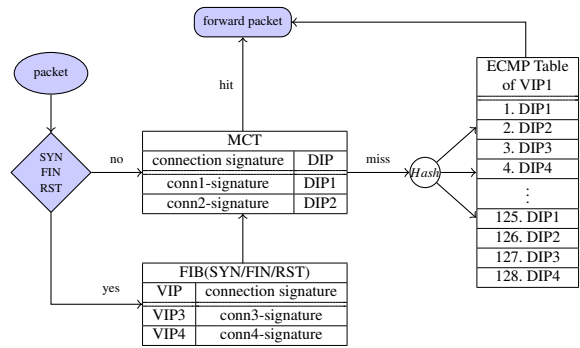

\par Prism distinguishes between a connection key and a connection signature. The key is a unique identifier, such as the connection's TCP/IP 5-tuple. This key is too long to be used in Prism's hardware tables (in particular for IPv6), and in the learning tables that capture information on the fly from SYN/FIN packets. Hence, Prism hashes the connection key to a shorter connection signature.

\par Prism maintains the association between connection signatures and DIPs using three tables:
\begin{enumerate}
\item  {\bf The SCT} (Software Connection Table)
\end{enumerate}
The SCT maintains the association between each connection signature and the DIP to which this connection has been mapped. This association maintains information about the SYN packet arrival time, and about other relevant parameters (see Section~\ref{sec:sct}). 

\begin{enumerate}[resume]
    \item {\bf The ECMP Table}
\end{enumerate}
The ECMP table uses standard ECMP logic in hardware, which forwards packets destined for the same logical address to a different "next hop", where next hop can be a port, a tunnel, an IP address, etc. Each VIP has its own ECMP table, which holds the VIP's DIP pool. Packets are mapped, based on their connection's signature, to one of the ECMP table entries. For example, the ECMP table shown in Figure~\ref{packet_flow_figure}  holds the pool \{DIP1, DIP2, DIP3, DIP4\} for VIP1. This specific table aims to balance the load between the four DIPs evenly, because 1/4 of the entries are allocated to each DIP.

\begin{enumerate}[resume]
    \item {\bf The MCT} (Migrated Connection Table) 
\end{enumerate}
The MCT maintains only migrated connections in hardware. A migrated connection is one for which the mapping between its signature and its DIP has been changed. In other words, the first packets of a migrated connection have been mapped by the ECMP table to a certain DIP, DIP1 say. But after a pool change, packets of this connection, if received by the ECMP table, would be forwarded to a different DIP, DIP2 say. To ensure that all the packets of this connection will continue to be forwarded to DIP1, the MCT maintains the association between this connection's signature and the original DIP, DIP1.

\par Figure~\ref{packet_flow_figure} depicts the flow chart of packets for a given VIP (VIP1). First, the MCT associated with this VIP is checked for a hit, which happens if the packet belongs to a migrated connection. If there is no MCT hit, the packet continues to the ECMP table, and finds there the DIP corresponding to its connection signature.

\section{Detailed Architecture}
\label{sec:detailed}

\subsection{The ECMP Table}
\label{sec:ecmp table}
Prism leverages programmable switch hardware logic, which uses generic match-action tables \cite{bosshart2013forwarding,jose2015compiling}. In this architecture, arriving packets are mapped to a specific table and are matched with an action. Standard switches use this logic for the implementation of ECMP-based diverse routing, where an ECMP table is chosen for each packet according to the packet's destination IP address. The table contains multiple equal cost forwarding path entries (also referred to as "bins"), and one of them is chosen for the considered packet using a hash function that takes into account other fields in the IP header, such as the source IP address and the port numbers.

\par In Prism, each ECMP table corresponds to a VIP, and each entry in the table contains the identity of a DIP. This identity could be the DIP's MAC address, but in order to reduce hardware cost, it actually contains a short "server ID" number. The packet takes this number and finds the actual DIP's address in another table.

\par An example of an ECMP table is shown in Figure~\ref{packet_flow_figure}. This table contains 128 entries and it serves a certain VIP (VIP1). This does not imply that there are 128 different physical servers (DIPs) for VIP1. Rather, these entries can be populated with any number (up to 128) of DIPs. By inserting the same DIP into multiple bins, Prism can enforce weighted load balancing among the various DIPs. In the example of Figure~\ref{packet_flow_figure}, there are only 4 DIPs and we want to balance the load between them evenly. Hence, the identities of these DIPs are equally populated among the 128 bins. An example of a non-uniform load balancing is assigning 1/3 of the bins to DIP1, 1/3 to DIP2, 1/6 to DIP3 and 1/6 to DIP4.

\par Changes to the DIP pool happen when the datacenter operators wants to add a DIP, to gracefully take down a DIP, or to change the load balancing weights. These events are discussed in Section~\ref{sec:dip pool update}. 

\subsection{The SYN Learning Tables}
\label{sec:syn_learning}
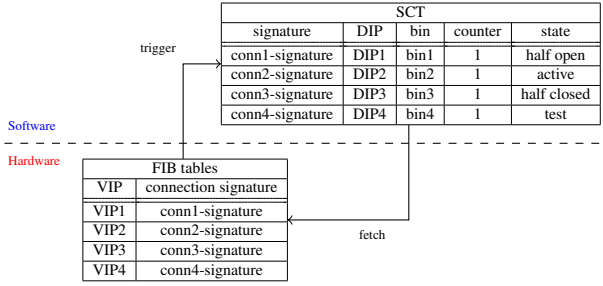
\begin{figure}[t]
\centering
\begin{tikzpicture}[node distance = 0.5cm, auto]

\node (fib) {
\scalebox{0.6}{
\begin{tabular}{ | c | c | }
  \hline
  \multicolumn{2}{|c|}{FIB tables} \\
  \hline
  VIP & connection signature \\
  \hhline{|=|=|}
   VIP1 & conn1-signature  \\
  \hline
  VIP2 & conn2-signature  \\
  \hline
  VIP3 & conn3-signature  \\ 
  \hline
  VIP4 & conn4-signature   \\ 
  \hline
\end{tabular}}};

\node [above = 0.2cm of fib, xshift=3cm] (ct){
\scalebox{0.6}{
\begin{tabular}{ | c | c | c | c | c |} 
  \hline
  \multicolumn{5}{|c|}{SCT} \\
  \hline
  signature & DIP & bin & counter & state\\
  \hhline{|=|=|=|=|=|}
  conn1-signature & DIP1  & bin1 & 1 & half open\\
  \hline
  conn2-signature & DIP2  & bin2 & 1 & active\\
  \hline
  conn3-signature & DIP3  & bin3 & 1 & half closed\\ 
  \hline
  conn4-signature & DIP4  & bin4 & 1 & test\\ 
  \hline
\end{tabular}}};
\node (cc) [yshift=0.25cm] at (ct.south){};
\node (dd) [xshift=-0.25cm] at (fib.east){};
\path [line] (cc) |- (dd);
\node [yshift=-0.2cm, right = 0.7cm of fib.east] {\tiny fetch};
\node (start) [yshift= 1.02cm, xshift = -2.5cm]  {};
\node (end) [right = 8cm of start]  {};
\path[draw, - ,dashed]  (start) node [above right = 2pt, blue] {\tiny Software} node [below right = 2pt, red] {\tiny Hardware} -- (end) ;
\node (aa) [yshift=-0.25cm] at (fib.north){};
\node (bb) [xshift=0.3cm] at (ct.west){};
\path [line] (aa) |- (bb);
\node [yshift= 0.2cm, left = 0.3cm of ct] {\tiny trigger};
\end{tikzpicture}
\caption{Connection insertion chart}
\label{entry_insertion_figure}
\vspace{-0.5cm}
\end{figure}

A TCP SYN sent by the client is the first packet of a new connection. When a SYN is received by Prism, two actions are needed. First, Prism needs to store the signature of the new connection in the SCT (Figure~\ref{entry_insertion_figure}). Second, this SYN has to be forwarded to the ECMP table and from there to the chosen DIP. 

\par Adding an entry to the SCT, which is a software table, takes a relatively long time. Thus, Prism uses a hardware table to learn the signatures of new SYNs in real-time, and only later, asynchronous to packet forwarding, adds the information about the signatures of new connections to the SCT. To learn the signature of new SYNs, Prism leverages the FIB hardware unit, which has been traditionally used for L2 MAC learning. After traversing the FIB(SYN) table, the SYN continues along the path shown in Figure~\ref{packet_flow_figure}.

\par To make sure that a SYN packet is not ignored by the server (e.g., because the server is not alive), Prism needs to know if the server also sends a SYN-ACK. LBs are usually divided into two categories with respect to the return traffic. In the first category, return traffic is also routed through the LB, while in the second category (direct server return) it does not. {\bf Since Prism is sitting on the switch,} we can enforce the return traffic to traverse it without encountering any routing penalty. We take advantage of this important property in order to capture in a FIB(SYN2) table the SYN ACK packets sent by the servers. Only when both SYNs are received, a connection in the SCT is considered active. Prism uses the SYN from the server to verify the validity of the connection. If a SYN2 is not received from the DIP within a timeout period, the entry is removed from the SCT.

\par The two FIB(SYN) tables, FIB(SYN1) and FIB(SYN2), are implemented as cuckoo hash tables \cite{pagh2004cuckoo,kirsch2009morecuckoo,fotakis2005spacecuckoo, levy2017flexible} in hardware. In cuckoo hashing, the table is split into multiple sections and each section has a different hash function. Suppose there are $N$ hash sections $S_1, \cdots ,S_N$ (N is usually $\geq$ 2). When a new connection signature $SIG_1$ is received, N hash functions are computed in parallel for the considered signature, and an index is determined for each section. Let the indexes be $i_1, \cdots ,i_N$. Then, $SIG_1$ is added to one (e.g., the first) section $j$ for which $S_j[i_j]$ is empty. If for every $j$, $S_j[i_j]$ is not empty, one of the sections $k$ is chosen randomly, and $SIG_1$ is swapped with the signature $SIG_2$ in $S_k[i_k]$. Then, the process is repeated for $SIG_2$. There are some tricks to ensure that the whole process is completed before a new signature is received \cite{kirsch2007using}, but this is outside the scope of this paper. 

\par Each FIB(SYN) table is read by the software periodically, e.g., once every 1 to 10 $m$s. The software then drains the table. This process is further discussed in Section~\ref{sec:dip pool update}, in the context of PCC.

\subsection{The FIN and RST Learning Tables}
A TCP FIN packet is sent by the client to indicate that it does not intend to send more data, and that it is willing to take down the connection when the server also finishes sending data. A similar packet is sent by the server when it does not intend to send more data on the connection. Prism uses a FIB(FIN1) table to learn about FIN packets sent by the clients, and a FIB(FIN2) table to learn about FIN packets sent by the server.

\par Both server and client may send RST packets to terminate a connection due to an unexpected error. To learn about RST packets sent by the client or the server, Prism uses FIB(RST1) and FIB(RST2) respectively.  

\subsection{The Software Connection Table (SCT)}
\label{sec:sct}
To track the state of active connections, and to track server pool changes while connections are alive, Prism maintains a state table for all active connections. As indicated in Section~\ref{sec:introduction}, a unique property of Prism compared to other hardware based LBs, is that this table is only maintained by the software. The software periodically reads the hardware FIB(SYN1) and FIB(SYN2) learning tables and adds into the SCT a state for each new connection with the following information:
\begin{enumerate}[label=(\alph*)]
  \item the connection signature;
  \item the DIP to which this connection signature is assigned;
  \item the ECMP bin index of the chosen DIP;
  \item the time when this connection signature is added to the SCT;
  \item the state of the connection (half-open, active, half-close etc);
  \item the number of connections mapped to this entry.
\end{enumerate}

\par To understand (f), it is important to note that two connections with two different keys (e.g., IPv4 5-tuples) might have the same signature. In such a case, a SYN for the second connection may arrive when there is already an SCT entry for the first. The probability for this anomaly is very low (see Section~\ref{sec:signatures}), but it must be taken into account in a system that receives millions of connections every second. To address this issue, an SCT entry actually corresponds to a signature, not to a connection, and Prism maintains a counter that indicates the number of different connections currently mapped to this entry. The way signatures are defined ensures that two connections with the same signatures are destined for the same VIP. Thus, they can share an SCT and an MCT entry. This is further discussed in Section \ref{sec:signatures}.

\par Connections that terminate should be removed from the SCT, and from the MCT if applicable. Thus, the software periodically reads the hardware FIB(FIN) and FIB(RST) tables, decrements the counter when a connection ends, and removes an entry when the reference count is 0.

\subsection{The Migrated Connection Table (MCT)}
MCT is a hardware table that maintains the mapping between signatures and DIPs of migrated connections. Recall that a migrated connection is a connection that was mapped to an ECMP bin, but the content of this bin has been changed later on. Such a bin is referred to as an "affected bin". Similarly to the FIB table, the MCT is implemented as a hardware-based cuckoo hash table.

\par As shown in Figure~\ref{packet_flow_figure}, every packet received by Prism is checked against the the MCT (a SYN/FIN/RST packet is first checked against the FIB and only then against the MCT). If the connection signature of this packet is not found in the MCT, then the content of the ECMP bin for this connection has not been changed since the arrival of the client's SYN. Hence, the original DIP for this connection can still be found in the ECMP bin associated with the packet's connection signature.

\begin{figure}[t!]
\centering
\begin{tikzpicture}[node distance = 2cm, auto] 
\node[rect,text width=2em, minimum height=2em] (none) {\tiny none};
\node [rect, right = 1.5cm of none] (sct) {\tiny connection\:is only\:in\:SCT};
\node [rect, right= 1.7cm of sct] (mct) {\tiny connection\:is\:in \:SCT\:and\:MCT};
\node [block, above = 1.3cm of mct] (remove) {\tiny{remove connection}};
\path[line] (none) -- node {\tiny receiving SYN1} (sct);
\path (sct.east) edge[yshift=2pt,bend left,->,dashed] node {\tiny migrated} (mct.west);
\path (mct.west) edge[yshift=-2pt,bend left,->,dotted] node {\tiny re-activated} (sct.east);
\path [line,bend right=90] (sct.north) -- ++(0,1.5) -- node[xshift=-2.3cm,yshift=-0.7cm] {\tiny receiving 2 FINs or RST} node[xshift=-2.2cm,yshift=-0.9cm] {\tiny or SYN2 timeout}  (remove.west);
\path (mct) edge[bend left,->,dashed] node[yshift=-0.2cm] {\tiny receiving 2 FINs or RST} (remove);
\path (mct) edge[bend right,->,dotted] node[right, yshift=0.05cm] {\tiny timeout} (remove);
\end{tikzpicture}
\caption{Prism's high level connection transition}
\label{fig:connection_life}
\vspace{-0.5cm}
\end{figure}
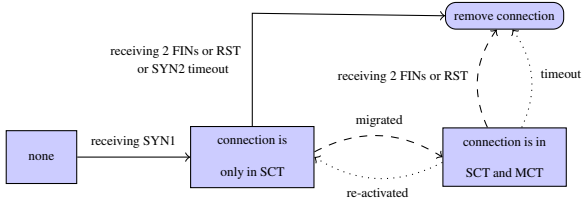
\subsection{Connection Transition Between Tables}
\label{sec:connection_trans}
Connections move across different states throughout their lifetime. Figure~\ref{fig:connection_life} shows the most important state transitions. Most connections make the following standard path: their SYN1 is received by the FIB(SYN1) table and their signature is inserted into the SCT; then, their SYN2 is received, and their state in the SCT is changed from half-open to open; then, their FINs are captured by the FIB(FIN) tables and they are removed from the SCT. Some connections make the dashed path upon DIP pool update and are migrated into the MCT. If a connection does not receive SYN2 after a short timeout (e.g., 50 $m$s), it is removed from the SCT, and also from the MCT if applicable. 

\par For a very short-lived connection, it is possible that Prism becomes aware of its FIN before its SYN. For example, this is the case for a connection that lasts only 5$m$s, if: (a) the FIB tables are read once every 10$m$s, (b) both SYNs are received a short time after FIB(SYN) is read, and (c) its FINs are received just before the FIB(FIN) is read. Prism puts such a connection in "only FIN received" SCT state. SCT entries in this state are pruned if matching SYN1/SYN2 packets are not received within a short timeout, which is equal to the FIB(SYN) polling interval.

\par Another uncommon case is when a connection is not terminated properly, i.e., Prism does not receive two FINs or a RST for it. To address this issue, consider first a connection that is also in the MCT. Tracking is done using a keep-alive bit in the MCT entry. This bit is set by every packet with the same signature. The software periodically checks this bit for each MCT signature. If it is set, the bit is cleared. If it is cleared, the signature is removed from the MCT and the SCT.

\par Now consider a connection that is not in the MCT. When such a connection remains for a relatively long time in the SCT compared to the expected connection time for the considered VIP, it is added into the MCT to track its activity. If it is found to be active, it is removed from the MCT and remains in the SCT; else, it is removed from both the SCT and MCT.

\section{DIP Pool Updates}
\label{sec:dip pool update}
When a DIP fails, all the connections assigned to it terminate prematurely. There is nothing the LB can do about such events, except replacing the identity of the failed DIP with the identity of another DIP in all relevant ECMP bins. This ensures that future connections will be forwarded to a live server. However, when the cloud operator wants to add more DIPs for a certain VIP, or to gracefully take down one or more DIPs, or to change the load distribution among the DIPs, active connections must not be affected \cite{govindan2016evolve}. This concept, called PCC, is guaranteed using the process of smooth DIP pool update, which we now describe. 

\par We first note that all the above mentioned events require a redistribution of the ECMP bins associated with the considered DIP. The update process is triggered by the software, and is divided into two parts: (a) copying the signatures whose bins content should be changed from the SCT to the MCT; (b) updating the relevant bins in the ECMP table.

\begin{figure}[t]
\centering
\begin{minipage}[b]{.4\textwidth}

\begin{tikzpicture}[node distance = 3cm, auto]
\node (ecmp_before) at (0,0){
\scalebox{0.6}{
\begin{tabular}{ | c | } 
  \hline
  ECMP Table\\
  \hhline{|=|}
  1. DIP1  \\
  \hline
  2. DIP2  \\
  \hline
  3. DIP3  \\ 
  \hline
  \rowcolor[HTML]{FF3E00} 4. DIP4  \\ 
  \hline
  $\vdots$ \\
  \hline
  125. DIP1  \\
  \hline
  126. DIP2  \\
  \hline
  127. DIP3  \\ 
  \hline
  \rowcolor[HTML]{FF3E00} 128. DIP4  \\ 
  \hline
\end{tabular}}};

\node (ecmp_after) at (2,0){
\scalebox{0.6}{
\begin{tabular}{ | c | } 
  \hline
  ECMP Table \\
  \hhline{|=|}
  1. DIP1  \\
  \hline
  2. DIP2  \\
  \hline
  3. DIP3  \\ 
  \hline
  \rowcolor[HTML]{08B000} 4. DIP1  \\ 
  \hline
  $\vdots$ \\
  \hline
  125. DIP1  \\
  \hline
  126. DIP2  \\
  \hline
  127. DIP3  \\ 
  \hline
  \rowcolor[HTML]{08B000} 128. DIP2  \\ 
  \hline
\end{tabular}}};
\draw[->,thick] (ecmp_before) -- node[] {\tiny update} (ecmp_after);

\node (a) [below right = 0.05cm of ecmp_before] {\small (a)};
\node (b) [right = 3.3cm of a] {\small (b)};

\node (ecmpBefore) at (4,0){
\scalebox{0.6}{
\begin{tabular}{ | c | } 
  \hline
  ECMP Table \\
  \hhline{|=|}
  1. DIP1  \\
  \hline
  2. DIP2  \\
  \hline
  3. DIP3  \\
  \hline
  4. DIP4 \\
  \hline
  \rowcolor[HTML]{33B5FF} 5. DIP1  \\
  \hline
  $\vdots$ \\
  \hline
  \rowcolor[HTML]{33B5FF} 10. DIP2  \\ 
  \hline
  $\vdots$ \\
  \hline
  128. DIP4  \\ 
  \hline
\end{tabular}}};

\node (ecmpAfter) at (6,0){
\scalebox{0.6}{
\begin{tabular}{ | c | } 
  \hline
  ECMP Table \\
  \hhline{|=|}
  1. DIP1  \\
  \hline
  2. DIP2  \\
  \hline
  3. DIP3  \\
  \hline
  4. DIP4 \\
  \hline
  \rowcolor[HTML]{33D1FF} 5. DIP5  \\
  \hline
  $\vdots$ \\
  \hline
  \rowcolor[HTML]{33D1FF} 10. DIP5  \\ 
  \hline
  $\vdots$ \\ 
  \hline
  128. DIP4  \\ 
  \hline
\end{tabular}}};
\draw[->,thick] (ecmpBefore) -- node[] {\tiny update} (ecmpAfter);
\end{tikzpicture}
\caption{ \centering (a) DIP4 is gracefully taken down; (b) adding a new DIP (DIP5)}
\vspace{-0.5cm}
\label{fig:ecmp_update_fail}

\end{minipage}
\end{figure}
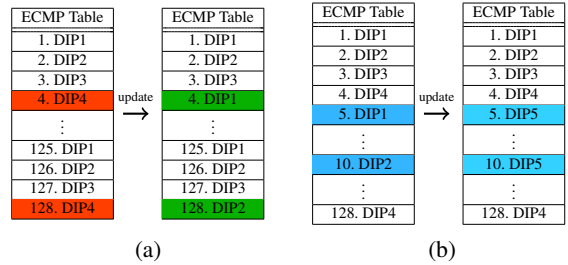

\par To understand (b), suppose that we have 128 bins for a given VIP, and these bins are equally populated with the identities of 4 servers: DIP1 -- DIP4. Suppose also that we want to take down DIP4 for maintenance. Thus, we need to copy into the MCT all the signatures that are currently mapped to bins that contains the identity of DIP4, and then to populate these bins with the identities of DIP1 -- DIP3. This is illustrated by Figure~\ref{fig:ecmp_update_fail}(a).

\par Next, suppose that we want to add a new DIP, DIP5, to the pool for a VIP, while maintaining statistical equal load among all DIPs. Thus, we need to have 128/5 bins assigned to DIP1 -- DIP5, instead of 128/4 bins assigned to DIP1 -- DIP4. To this end, 128/4-128/5 = 6 bins should be chosen from DIP1 -- DIP4, and assigned to DIP5. This is illustrated by Figure~\ref{fig:ecmp_update_fail}(b): every bin $i$ for which i mod 5 = 0 is populated with the identity of DIP5. For each chosen bin, all the signatures mapped to this bin are copied into the MCT, and only then will the bin be populated with the identity of DIP5.



\par Changing the load distribution of a certain VIP among the existing bins is performed in a similar way. Suppose that the datacenter operator wants to double the load imposed on DIP1 and DIP2 compared to DIP3 and DIP4. Thus, the load distribution should be 1/3, 1/3, 1/6, 1/6. Assuming that before this change the bins are populated evenly, Prism chooses 128/4-128/6 = 10 bins that are currently assigned to DIP3, and the same number of bins that are currently assigned to DIP4, and divides them evenly between DIP1 and DIP2. 


\par
\begin{figure}[t]
    \centering
\begin{tikzpicture}
\draw (0,0) -- (6.5,0);
\foreach \x in {0.5,3.8,5,6}
\draw(\x cm,3pt) -- (\x cm, -3pt);
\draw (0.5,0) node[below=3pt] {${t}_{1}^{\mbox{\em \tiny {start}}}$};
\draw (3.8,0) node[below=3pt] {${t}_{1}^{\mbox{\em \tiny {end}}}$};
\draw (5,0) node[below=3pt] {${t}_{2}^{\mbox{\em \tiny {start}}}$};
\draw (6,0) node[below=3pt] {${t}_{2}^{\mbox{\em {\tiny {end}}}}$};

\foreach \x in {1.5,2,2.5}
\draw[->, thick] (\x,1.2) -- (\x,0.2); 
\draw (2,1.2) node[above=3pt] {$\mbox{\em new SYNs}$};

\draw [decorate,decoration={brace,mirror,amplitude=5pt}] (0.5,-0.7) -- node [below = 5pt] {$\mbox{\em first round}$} (3.8,-0.7);

\draw [decorate,decoration={brace,mirror,amplitude=5pt}] (5,-0.7) -- node [below = 5pt] {$\mbox{\em second round}$} (6,-0.7); 

\draw (5.5,1.2) node[above=3pt] {$\mbox{\em no new SYNs}$};
\draw[->, dashed, thick] (5.5,1.2) -- (5.5,0.2); 
\end{tikzpicture}
    \caption{Ensuring PCC during connection migration; in this example two migration rounds are sufficient, because the second round is too short for new SYNs to arrive.}
    \label{fig:migration timeline}
\end{figure}
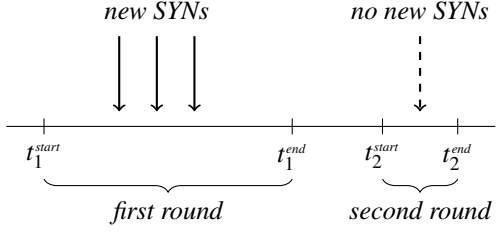

 In all the above cases, \textbf{for each affected bin}, Prism first copies to the MCT the signatures that have been mapped into this bin and only then replaces the bin's content. To perform this process without breaking PCC, the following race condition should be addressed (Figure~\ref{fig:migration timeline}). Suppose that at ${t}_{1}^{\mbox{\em \tiny {start}}}$ Prism starts copying signatures of a certain bin from the SCT to the MCT. Suppose that there are $C_1$ such signatures, and copying them takes $T_1$ $m$s. During this time, new connections continue to arrive and their signatures are recorded by FIB(SYN1). Some of these signatures, say $C_2$, are mapped to the considered bin. All these signatures must also be inserted into the MCT, because their status is similar to that of the first $C_1$ affected signatures. Thus, when Prism finishes copying the first $C_1$ affected signatures from the SCT to the MCT, at ${t}_{1}^{\mbox{\em \tiny {end}}}$, it copies all new signatures from the FIB(SYN1) table to the SCT, and starts the second migration round, at time ${t}_{2}^{\mbox{\em \tiny {start}}}$. During this round, the $C_2$ newly affected signatures of the considered bin are copied from the SCT to the MCT. As shown later, it is expected that $C_2 \ll C_1$ holds. Hence, the second migration round, which takes $T_2$ $m$s, is expected to be much shorter compared to the first one; namely $T_2\ll T_1$ holds. Again, new signatures for the considered bin may be received during $T_2$. Their number, $C_3$, is expected to be less than 1. If this is the case, Prism stops the MCT update process for the considered bin, and can move to another affected bin, if such exists. Otherwise, another round can be initiated. The above process is summarized by Algorithm~\ref{alg:iterative migration}.

\SetKwInput{KwInput}{Input}                
\SetKwInput{KwOutput}{Output}              
\SetKwInput{KwStart}{Start}              
\begin{algorithm}
\SetAlgoLined
\KwInput{\\
$T_{\mbox{\em \tiny limit}}$: Tolerable time for updating the considered bin}
\KwStart{}
 read FIB(SYN1) and insert the new signatures into the SCT;\\
 build the initial set $C_1$ of the signatures of the considered bin;\\
 $i \leftarrow 1$;\\
 $Last \leftarrow FALSE$;\\
 \While{$C_i > 0$ \mbox{\em and} not $Last$}{
   \If{\mbox{\em There is no time for an additional $(i+1)$'th round}}{
      configure FIB(SYN1) to forward to the software new SYNs for the considered bin;\\
      $Last \leftarrow TRUE$;\\
   }
   copy the $C_i$ affected signatures into the MCT;\\
   ${i \leftarrow i+1}$;\\
   \If{not Last}{
    read FIB(SYN1) and insert new signatures into the SCT;\\
    build $C_i$;\\
   }
 }
 \caption{Iterative Migration of Bin Signatures}
 \label{alg:iterative migration}
\end{algorithm}

\par Algorithm~\ref{alg:iterative migration} starts the $i$'th round, where $i$>1, only if $C_{i} > 0$. When it starts the $i$'th round, it decides whether this will be the last round. The decision is made using a parameter $T_{\mbox{\em \tiny limit}}$ that indicates the maximum time after which the algorithm must terminate in order not to delay the pool update too much. If the algorithm decides before starting the $i$'th round that there is no time for the $(i+1)$'th round, it configures FIB(SYN1) to forward to the software all the SYNs received for the considered bin during the $i$'th round. These SYNs, whose expected number is shown later to be very close to 0, are added directly to the SCT. The implication is that the connections that send these SYNs are slightly delayed, but they are not lost.

\par To analyze the algorithm, let $\rho$ be the rate of copying signatures from SCT to MCT, and $\delta$ be the rate of receiving new signatures for the considered bin. The time it takes for the first MCT update round is $T_1=\frac{C_1}{\rho}$. The time for reading the FIB(SYN) tables is much faster compared to the time for copying a signature from the SCT to the MCT: 0.1$\mu$s (10M per second) vs. 0.04$m$s (25,000 per second). The number of signatures arriving for the considered bin during the first round is ${C_2=\delta \cdot \frac{C_1}{\rho}}$. Thus, the expected length of the $i$'th update round is ${T_i=(\frac{\delta}{\rho})^{i-1} \cdot \frac{C_1}{\rho}}$, and the number of signatures arriving for the considered bin during this time interval is ${C_{i+1}=(\frac{\delta}{\rho})^{i} \cdot C_1}$.

\par The condition for ensuring that $C_{i+1} \ll C_i$ is $\frac{\delta}{\rho} < 1$ . This condition is easily satisfied for all realistic scenarios, as shown below. The $i$'th round is the last one if: ${(\frac{\delta}{\rho})^{i} \cdot C_1 < 1}$ or if the total running time including the expected running time of the $i$th round ($T_{i}$) is greater than $T_{\mbox{\em \tiny limit}}$; i.e., $T_{\mbox{\em \tiny total}} + E(T_i) > T_{\mbox{\em \tiny limit}}$. Thus, the expected number of rounds, unless $T_{\mbox{\em \tiny limit}}$ is reached, is ${i = \frac{\log C_1}{\log \frac{\rho}{\delta}}}$.

\par As an example, consider the following realistic numbers:
\begin{itemize}
  \item Writing 25 signatures into the MCT takes 1$m$s.
  \item The LB receives 1M new connections every second.
  \item The LB handles 100M concurrent active connections.
  \item There are 100,000 ECMP bins. Thus, each bin holds, on the average, 1/100,000 = 1,000 of the active connections.
\end{itemize}

With these numbers, $C_1 \approx 1,000$, $\delta$ = 10 connections per second and $\rho$ = 25,000 signatures per second. Thus, it takes $T_1$ = 40$m$s to insert the first $C_1$ affected signatures into the MCT during the first round. Meanwhile, $C_2$ $\approx$ 0.4 new signatures arrive to the considered bin and are recorded by FIB(SYN1). During the first round (40$m$s), 40,000 new signatures were recorded by the FIB(SYN1). Thus, the whole FIB(SYN1) is read in 4$m$s. If a signature for the considered bin was recorded, a new round is initiated. In the second round, the signature is inserted into the MCT in $T_2$ = 40$\mu$s. During this time, no new signatures are expected to arrive to the considered bin. 

\par Generally, under realistic scenarios, $T_{\mbox{\em \tiny limit}}$ is long enough for the algorithm to finish without deflecting any SYNs to the software. In the example above, the total running time of the algorithm is 44.04$m$s. If instead of 1,000 signatures, the considered bin starts with 10,000 signatures (x10 of the average), then the algorithm would need a second round to copy $C_2 \approx$ 4 signatures, and the total execution time is 440.16$m$s.

\ifLongversion
\begin{figure*}[ht!]
\centering
\begin{minipage}[t]{.4\textwidth}
\scalebox{0.45}{\input{simulation_figures/frequent_updates_10000_dips_2_sec_avg.pgf}}
\caption*{\small \textmd{(a) average connection length = 2s}}
\end{minipage}\qquad\qquad
\begin{minipage}[t]{.4\textwidth}
\scalebox{0.45}{\input{simulation_figures/frequent_updates_10000_dips_5_sec_avg.pgf}}
\caption*{\small \textmd{(b) average connection length = 5s}}
\end{minipage}
\caption{\centering The percentage of connections stored in the MCT throughout system running time as a function the number of updates per minute}
\label{fig:frequent_pool_update}
\end{figure*}
\else
\begin{figure}[t]
    \centering
    \scalebox{0.45}{\input{simulation_figures/frequent_updates_10000_dips_5_sec_avg.pgf}}
    \caption{\centering The percentage of connections stored in the MCT throughout system running time as a function of the number of DIP updates per minute}
    \label{fig:frequent_pool_update}
\end{figure}
\fi

\par We ran simulations to measure the load imposed on the MCT by Algorithm~\ref{alg:iterative migration} as a function of the pool update frequency.  Figure~\ref{fig:frequent_pool_update} shows the percentage of connections stored in the MCT, for several pool update rates. These experiments were conducted for 100 VIPs, each with 100 DIPs, serving roughly the same number of active connections. There are 10,000 active ECMP bins, and each DIP resides in one bin. In each experiment, a random DIP is taken down for maintenance, and a new DIP takes its place. Each experiment is repeated 5 times with a different seed, and a point in the graph is created for the average of all these runs. The maximum pool change rate considered in these simulations is 120 per minute, namely, two servers are taken down every second, on the average. 

\par All experiments start with an empty MCT. The SYN arrival rate for each VIP is  10,000 per second (1M connections per second for all the VIPs), and each connection lasts between 1 to 10 seconds. Connection lifetimes are uniformly distributed, with an average of \ifLongversion 2 (Figure~\ref{fig:frequent_pool_update}(a)) or 5 (Figure~\ref{fig:frequent_pool_update}(b)) seconds\else 5 seconds.\fi We can see that  in the worst case, with 2 pool updates per second, only 0.06\% of the connections are added to the MCT.

\par During the above experiment, we also measured the number of migrated connections per pool update. The results are shown in Table~\ref{tab:migrations}. The implication of these results is that with MCT writing rate of 25 signatures per 1$m$s, as in our prototype, the migration process (Algorithm~\ref{alg:iterative migration}) lasts, on the average, 10$m$s or 22$m$s, when the average connection length is 5 second or 2 second, respectively.

\begin{table}
\newcolumntype{C}[1]{>{\centering\arraybackslash}m{#1}}
    \centering
    \begin{tabular}{ | c | c | c | }
      \hline
     \multirow{3}{1cm}{\centering updates per second} &  \multicolumn{2}{C{5.6cm}|}{average number of migrated signatures per update} \\
     \cline{2-3}
        &  \multirow{2}{2.8cm}{\centering average connection length = 2 second} &  \multirow{2}{2.8cm}{\centering average connection length = 5 second}\\
       & &\\
      \hline
      15 & 249.66 & 543.74 \\
      \hline
      30 & 250.09 & 545.56 \\
      \hline
      60 & 249.41 & 546.01 \\
      \hline
      120 & 249.12 & 545.31 \\
      \hline
    \end{tabular}
    \caption{\centering Average number of migrated signatures per update}
    \label{tab:migrations}
\end{table}

\par
\begin{figure}[t]
\scalebox{0.45}{\input{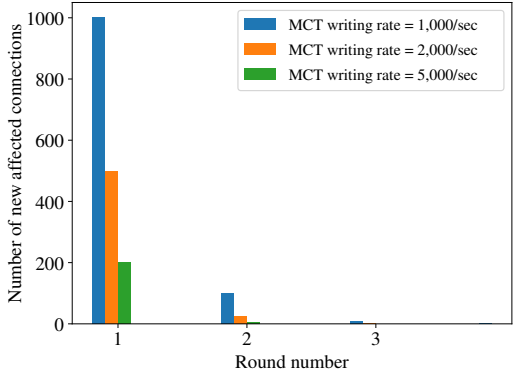}}
\caption{\centering The effect of the MCT writing rate ($\rho$) on the number of rounds of Algorithm~\ref{alg:iterative migration} }
\label{fig:ecmp_size_graph}
\end{figure}

 Next, we analyze the effect of burst arrival of SYNs. To this end, we ran simulations for the case where the migrated bin has 10,000 affected signatures, namely, 200 times more than the average bin occupancy, the length of the ECMP table of the considered VIP is 100, and the arrival rate of SYNs for the considered VIP is 10,000 per second. This means that there are $\delta =$ 100 SYNs per second per bin, which is 10 times more than the average rate when all 100,000 bins are active. We generated connections with random signatures and count the number of SYNs matching the considered bin throughout the iterative update algorithm. Figure~\ref{fig:ecmp_size_graph} shows the number of new affected signatures received during each round of the update algorithm as a function of the MCT writing rate ($\rho$). The presented values are the average of 2,000 independent runs of the algorithm. We can see that even for low values of $\rho$, all runs required no more than 3 rounds of Algorithm~\ref{alg:iterative migration} to finish the migration.

\par The decision on which bins to choose for the update process should not be made randomly. Appendix~\ref{sec:choosing bins} presents several methods and discusses their performance.

\section{Signatures and Collisions}
\label{sec:signatures}
\subsection{The Various Collision Scenarios}
\label{sec:collision scenarios}
\begin{figure}[h!]
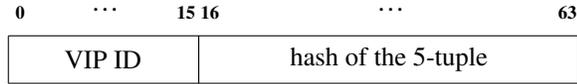

\begin{center}
\newcommand\bithead[2]{\bitbox[]{#1}{\scriptsize\bfseries #2}}
\begin{bytefield}[bitwidth=0.9em]{24}
  \bithead{1}{0} &
  \bithead{6}{. . .} &
  \bithead{1}{15} &
  \bithead{1}{16} &
  \bithead{14}{. . .} &
  \bithead{1}{63}\\
  \bitbox{8}{VIP ID} &
  \bitbox{16}{hash of the 5-tuple}\\
\end{bytefield}
\end{center}
\vspace{-0.5cm}
\caption{64-bit signature used in our prototype}
\label{fig:signature}
\vspace{-0.2cm}
\end{figure}

In this section we discuss the connection signature as used in our prototype, for a system with up to $2^{16}$ VIPs. This signature can be modified, of course, to address different requirements. The signature consists of two parts (Figure~\ref{fig:signature}): a 16-bit VIP ID, and a 48-bit hash of the connection key (5-tuple in IPv4). The translation from the destination IP address into a VIP ID can be performed in several ways. In our prototype we use a translation table, implemented as a cuckoo hash table.

\par Using signatures introduces collisions. We actually may have two collisions types. The first is when two different connections, with different 5-tuples, are hashed to different
signatures, but these two signatures are mapped to the same  cuckoo hash table entry. 
The second is when two different connections, with different 5-tuple, happen to have the same
signature. The first type of collisions is resolved using standard cuckoo
hashing techniques \cite{kirsch2007using,pagh2004cuckoo,bosshart2013forwarding}. Hence, from now on focus on collisions from the second type.

\par Following our design, when two different connections have the same signature, they are destined for the same VIP. Suppose that two SYNs with the same signature, $\text{syn}'$ and $\text{syn}''$, are recorded by FIB(SYN1), and let $\text{syn}'$ be the first. When $\text{syn}''$ is recorded by FIB(SYN1), there are three options:
\begin{enumerate}[label=(\alph*)]
\item the signature of $\text{syn}'$ has already been written into the SCT, and it has not been migrated into the MCT;
\item the signature of $\text{syn}'$ has already been written into the SCT, and con1 has been migrated into the MCT;
\item $\text{syn}'$ has not been written yet into the SCT; namely, $\text{syn}''$ is matched on FIB(SYN1) while it already contains the signature of $\text{syn}'$.
\end{enumerate}
\par In option (a), when Prism copies signatures from FIB(SYN1) into the SCT it realizes that a signature identical to that of $\text{syn}''$ already exists in the SCT. If two SYNs with the same signature are received within a short (configurable) time $\Delta$, e.g., $\Delta$ = 5 second, Prism considers them as belonging to the same connection. The reason is that SYN retransmissions is a much more likely event than two SYNs of different connections but with the same signature being received very close to each other. Thus, if $\text{syn}''$ is received less than $\Delta$ seconds after $\text{syn}'$, it is considered as belonging to the same connection. If not, they are considered as belonging to different connections, con1 and con2, and the counter associated with this SCT entry is incremented. In addition, the receiving time of $\text{syn}''$ replaces the receiving time of $\text{syn}'$ in their shared SCT entry.

\par In option (b), if $\text{syn}''$ is received less than $\Delta$ seconds after $\text{syn}'$, it is ignored. Otherwise, we distinguish between the case where the VIP pool update that migrated the signature of $\text{syn}'$ (and $\text{syn}''$) is still in progress, and the case where it has ended. In the first case, the SCT counter of this signature is incremented, but there is no need to do any change in the MCT. In the second case, $\text{syn}''$ and all following packets of its connection are routed according to the information in the MCT. When the SCT is aware of $\text{syn}''$, it updates the reference counter and the timestamp of the corresponding SCT entry.

\par The third option is very rare, because it requires that two SYNs of different connections will have the same signature and will be received within the same 10$m$s FIB periodic polling interval. The probability for this is practically 0. But in any case, it is a special case of $\text{syn}''$ being received less than $\Delta$ seconds after $\text{syn}'$, which means that it is ignored by Prism.

\subsection{The Rare Event of Losing A Connection}
\label{sec:rare collision}
As discussed in Section~\ref{sec:collision scenarios}, when two SYNs with the same signature are received less than $\Delta$ seconds from each other, they are considered as belonging to the same connection. This working assumption might be incorrect in certain rare cases. Suppose that $\text{syn}'$ and $\text{syn}''$ belong to different connections, con1 and con2 respectively, but they are received less than $\Delta$ second from each other. Thus, Prism incorrectly considers $\text{syn}''$ as belonging to con1 as well. We can now distinguish between two cases.

\par The first case is that the considered signature is not migrated into the MCT.
Without loss of generality suppose that con1 ends before con2. When the two FINs of con1 are recorded, the signature is removed from the SCT. The packets of con2 continue to arrive and to be forwarded correctly, according to the associated ECMP bin. Later, when con2 terminates, the software  will be informed about  two FINs for a signature that does not exist in the SCT. These FINs are ignored after a short timeout (see the discussion about the "only FINs received" SCT state in Section ~\ref{sec:connection_trans}), and everything returns to normal.

\par The second case is that the bin of these signatures is affected by a pool change before both connections terminate. If this happens while both connections are alive, the signature of these connections is migrated into the MCT and both connections will work correctly. But when con1 finishes, the MCT entry is removed, and the packets of con2 are directed to the wrong DIP. Thus, this connection will be broken. If this happens when con2 is alive and con1 is not, the signature is not migrated into the MCT and con2 is broken. 

\par To summarize, Prism loses a connection if two connections with the same signature are received within $\Delta$ seconds from each other and this signature is migrated into the MCT. The probability for this combination of events is very small, as shown in the following mathematical analysis.
 
\subsection{Mathematical Analysis}

\begin{table}[t]
    \centering
    \begin{tabular}{|c|c|c|c|}
    \hline
     \backslashbox{s}{m}  &  10 &  30 & 60 \\
    \hline
     10,000  & 6.39 & 2.13 & 1.07 \\
     \hline
     50,000  & 31.9 & 10.65 & 5.32 \\
     \hline
     100,000 & 63.89 & 21.3 & 10.65 \\
     \hline
    \end{tabular}
    \caption{\centering The expected time (in hours) until a connection is accidentally dropped, as a function of DIP migration rate (m) and the number of DIPs (s), with 1 million SYNs arriving to 1 VIP}
    \label{tab:time unti broken}
\vspace{-0.8cm}
\end{table}

To calculate the probability of receiving two SYNs with the same signature within $\Delta$ seconds, we consider this is an instance of the "birthday problem" \cite{wagner2002generalized}. The probability of at least one such event to happen is defined as: 
$$ p(n;d) = 1- \prod_{k=1}^{n-1} (1-\frac{k}{d}) \approx 1 - e^{-\frac{n^2}{2d}},
$$
where $n$ $(n \le d)$ is the number of SYNs arriving during a $\Delta$--second interval, i.e., $n = \Delta \cdot \delta$, and $d$ is the total number of possible signatures, i.e., $d = 2^{48}$  signatures per VIP. Consider a connection arrival rate of 1M per second, and the worst case where there is only 1 VIP in the system. The probability that a collision happens within a window of $\Delta$ = 5s, during which $5 \cdot 10^6$ connections are received, is: $$p(5,000,000;2^{48}) \approx 1 - e^{-\frac{5,000,000^2}{2^{49}}} \approx 0.0434.$$
Thus, the expected time elapsed until at least one event like this happens is:
$$
T_{\mbox{\em {\tiny between collisions}}} = \frac{1}{p(5,000,000;2^{48})} \approx \text{23 seconds.}
$$
This implies that with the considered numbers, a SYN would be mistakenly ignored every 23 seconds. As indicated in Section~\ref{sec:rare collision}, in order for this ignored SYN to break a connection, its signature should experience a migration. The probability for a signature to be migrated during its lifetime is given by:
$$p_{\mbox{\em \tiny migrated}}(m;s) = 1- (\frac{s-1}{s})^m ,$$
where $m$ is the number of DIP migrations that happen during the expected lifetime of a signature, and $s$ is the number of DIPs in the system. Table~\ref{tab:time unti broken} presents the expected time (in hours) until Prism is expected to accidentally drop a connection, namely, [ $\frac{1}{p_{\mbox{\em \tiny migrated}}}\cdot 23$  seconds], for several combinations of m and s. Note that the analysis and the results assume the worst case scenario, where all traffic goes to a single VIP. With more VIPs in the system, connection drops is much more rare.




\ifLongversion

\fi

\subsection{Removing Dead Connections}
SYN packets may be retransmitted, i.e., transmitted more than once for the same connection. As explained in Section~\ref{sec:signatures}, Prism considers SYNs with the same signature received during a short period of time ($\Delta$) as belonging to the same connection. 

\par However, it is possible that two SYNs of the same connection, con1, are sent during a period of time greater than $\Delta$. If both SYNs are responded by the server, Prism assumes that there are two connections with the same signature. Hence, the counter associated with this signature in the SCT is 2.  After FINs are sent for con1, Prism decrements the counter to 1, but no more FINs will be received for this signature, and the connection will remain in the SCT. 

\par Another scenario, even more common, which also results in a dead connection, is when the client fails before it sends a FIN for its connection. 

\par To solve this problem, Prism needs to track the status of "suspected signatures", namely, signatures that are marked "active" in the SCT for a long time period relatively to the expected lifetime of connections for the considered VIP. Since the software is not aware of connection activity status, Prism copies into the MCT suspected signatures that are only stored in the SCT. In the MCT, each entry is associated with a keep-alive bit. This bit is set by the hardware whenever a packet hits the entry. The software periodically reads the MCT entries of suspected signatures to check their activity bit. If the activity bit is set, the suspected signature is considered active, and is removed from the MCT. Otherwise, the software removes the signature from both the SCT and the MCT. This algorithm requires the SCT to maintain statistics about the expected connection lifetime for each VIP. A similar mechanism can be used if the reverse traffic is not routed through Prism, but this is outside the scope of this paper.


\subsection{SYN Flood Attacks}
In a SYN flood attack, high rate of fake SYNs are sent to the targeted server, without completing the TCP 3-way handshake. Consequently, the server is presented with many half-opened connections, and it risks not being able to receive legitimate  connections. This is a very effective attack, which can be performed distributedly,  using a spoofed source IP addresses.

\par Unlike a server, which is aware of half-opened connections, Prism thinks that these connections are normal, because it receives both the client's and server's SYNs. Thus, only after a half-opened connection stays too long in the SCT, it will be tested for activity and then pruned from the LB. This processes could be too costly when the attack rate is very high. 

SilkRoad addresses this problem by associating a rate-limiter to a VIP, to
detect and drop excessive traffic\cite{miao2017silkroad}. This approach can also
work with Prism, but in both cases the LB does not distinguish between real SYNs
and fake SYNs, and drop them together. To solve this problem, we have developed a hardware-based approach, which can distinguish between real and fake SYNs even if the rate of the fake SYNs is 10-time larger than the rate of real SYNs (i.e., $\approx$ 10 million per second). The description of this
extension is beyond the scope of this paper, and is given in \cite{prism_syn_attack}.

\section{Proof of Concept Prototype}
\label{sec:prototype}
\subsection{General Description}
We performed measurements and timing analysis on a Proof of Concept (POC) implementation using P4 \cite{bosshart2014p4} on top of a  Spectrum 2 \cite{spectrum} ASIC running SONiC \cite{sonic} on a x86 CPU (Broadwell, 8GB DRAM). The analysis focuses on three main areas of validation: the maximum learning rate for new connections, for which connection signatures are polled from the hardware to the software; programming new MCT entries from the software to the hardware as a result of DIP pool changes; and the average capacity of the hardware MCT under constant rate traffic conditions for up to 100 million active connections stored in the SCT.

\par The POC implementation is based on the hybrid target architecture. This model combines traditional switch functionality along with a multi-stage programmable pipeline. It provides all the benefits of built-in protocol handling, such as MAC learning and ARP handling, and standard network functions, such as L2/L3 forwarding, with a discrete pipeline. Switch programmability is exposed via an application sandbox, where plug-in points are provided to extend functionality, without requiring the programmer to re-write the legacy parser and forwarding functions in the pipeline. The POC implementation is deployed within a SONiC docker container. When configured and enabled, it seamlessly adds the load balancing capability to the standard set of SONiC supported network functions.

\begin{figure}[t!]
\scalebox{0.6}{\includegraphics[ trim = 3cm 5cm 5mm 5cm, clip,width=0.8\textwidth]{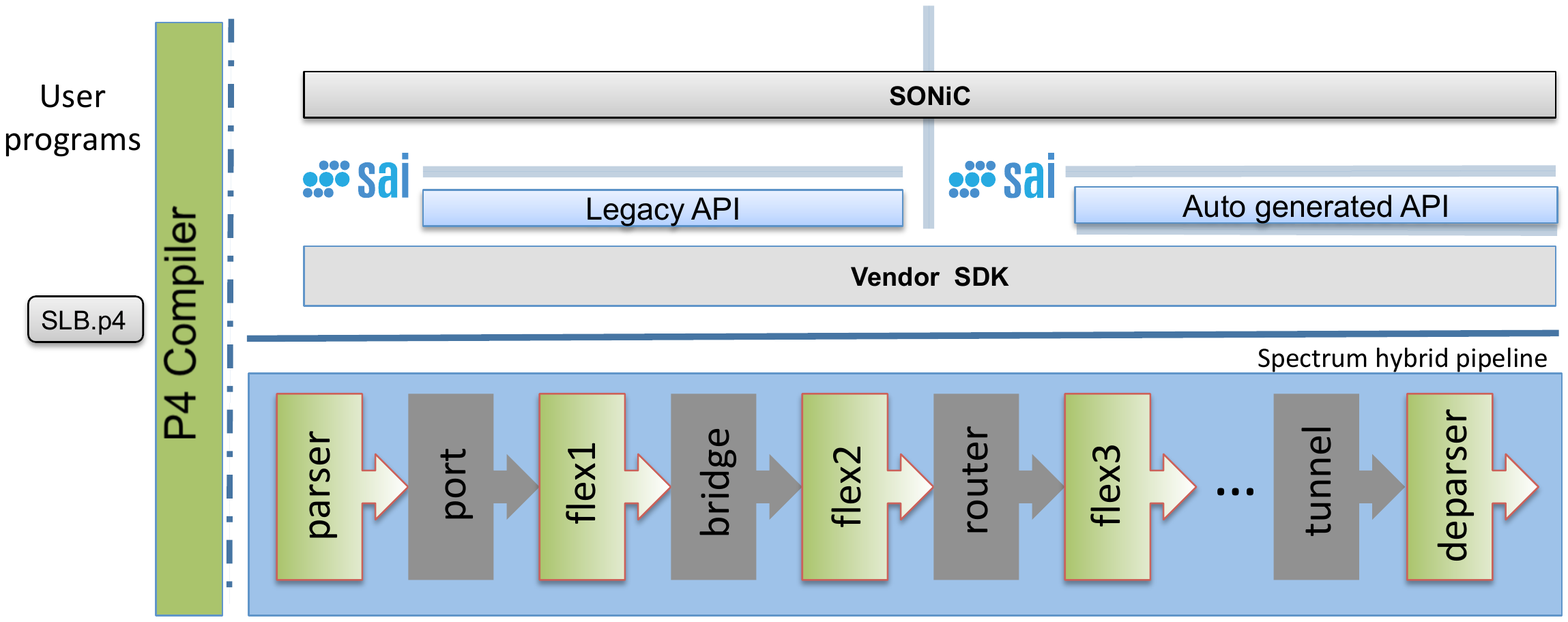}}
\caption{\centering Spectrum 2 hybrid programmable pipeline }
\label{fig:hybrid_pipeline}
\end{figure}

Figure \ref{fig:hybrid_pipeline} shows the packet processing pipeline stages. In the discrete pipeline, the green stages denote the flexible/programmable blocks where users can drop in extra functionality. Between them are grey stages that represent standard/legacy blocks that are pre-defined by the architecture. Our implementation leverages the P4 language and the Mellanox P4 compiler to extend the hybrid pipeline with the load balancing functionality. A description of our P4 implementation is presented in Appendix~\ref{app:p4_implementation}.

\subsection{POC Topology}
\begin{figure}[t]
\scalebox{0.7}{\includegraphics[scale = 2,trim = 5cm 4cm 5mm 5cm, clip,width=0.8\textwidth]{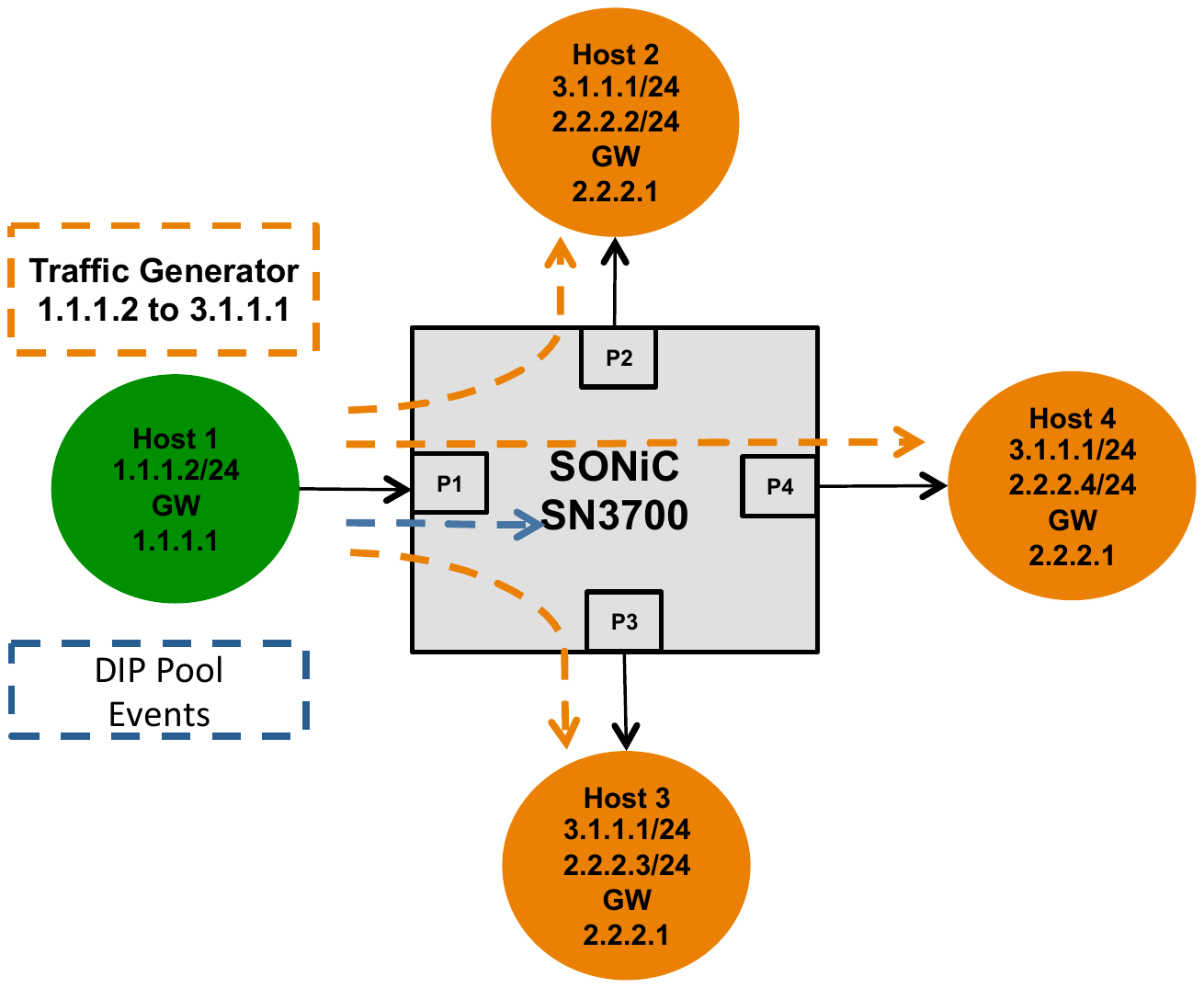}}
\caption{\centering POC testbed topology }
\label{fig:poc_topology}
\end{figure}

Our POC testbed topology is shown in Figure~\ref{fig:poc_topology}. On Host 1 is our traffic generator, which sends SYN/FIN packets according to a log normal distribution of connection. The packet generation is split into three phases: first, TCP SYN packets are generated, with the goal of filling the connection table close to the desired capacity. In the second phase, a constant rate (on average) of SYNs and FINs are generated to maintain the desired number of active connections. This equilibrium state is approximate, as the connection lifetimes are on purpose not of any fixed duration, in order to avoid artifacts due to discrete  connection start and termination intervals. It is during this phase that the DIP pool event manager sends a random add or remove DIP notification to the switch. The last phase of the packet generation is a graceful shutdown, where all SYN generation is halted and the remaining FINs are allowed to drain and close out the connections.

To further simplify the abstraction of DIPs, we configure 2 -- 4 receiving hosts to be the backend servers. Each host is configured with the IP address corresponding to the VIP which is the destination address used by to the traffic generator. We avoid the need to implement NAT for the server return path by assigning an internal secondary IP address and a static route back to the switch. In practice, we can split the 10K DIPs evenly between the 3 backend servers, each with its own secondary IP address. This testbed approach does not affect the measurements or scale we are attempting to obtain on the switch.

\vspace{-0.3cm}
\subsection{Learning Rate}
A crucial part of the implementation is the ability to rapidly learn new connections from the hardware, where the connection signature is calculated and stored in the hardware tables for the software to learn. It is important to recall that line rate packet forwarding is maintained since the software polling process is {\bf asynchronous}. No packets are held during the connection learning phase, and there is no added latency per connection. As indicated in Section~\ref{sec:syn_learning}, we leverage the hardware FIB mechanism for MAC learning, to store signatures of SYN packets, and notify the software. Each hardware notification contains a batch of new connection signatures, in order to make the polling and retrieval process efficient. On our prototype switch, the hardware FIB table is capable of holding up to 512K entries, which is later shown to be sufficient for the proposed scale.

\ifLongversion
\par
Our prototype uses the following configuration profile:
\begin{center}
\begin{tabular}{ | l | c | } 
\hline
Maximum number of active connections& 100M \\ 
\hline
Average connection lifetime & 1-100s \\ 
\hline
Connection arrival rate & 1M/s \\ 
\hline
Number of DIPs & 10,000 \\ 
\hline
Average number of connections per DIP & 10,000 \\ 
\hline
\end{tabular}
\end{center}
\fi

\par Our platform can read up to 10 million entries per second from the ASIC memory. The hardware can maintain, on average, 10,000 entries in the FIB table.  Thus, using a software polling rate of 10$m$s, and retrieving approximately 10,000 entries (SYNs) per poll, we can drain a very long burst of 1 million new connections per second. The CPU can maintain a full state for up to 100 million active connections.

\par
\begin{figure}[t]
\centering
\scalebox{0.36}{\includegraphics{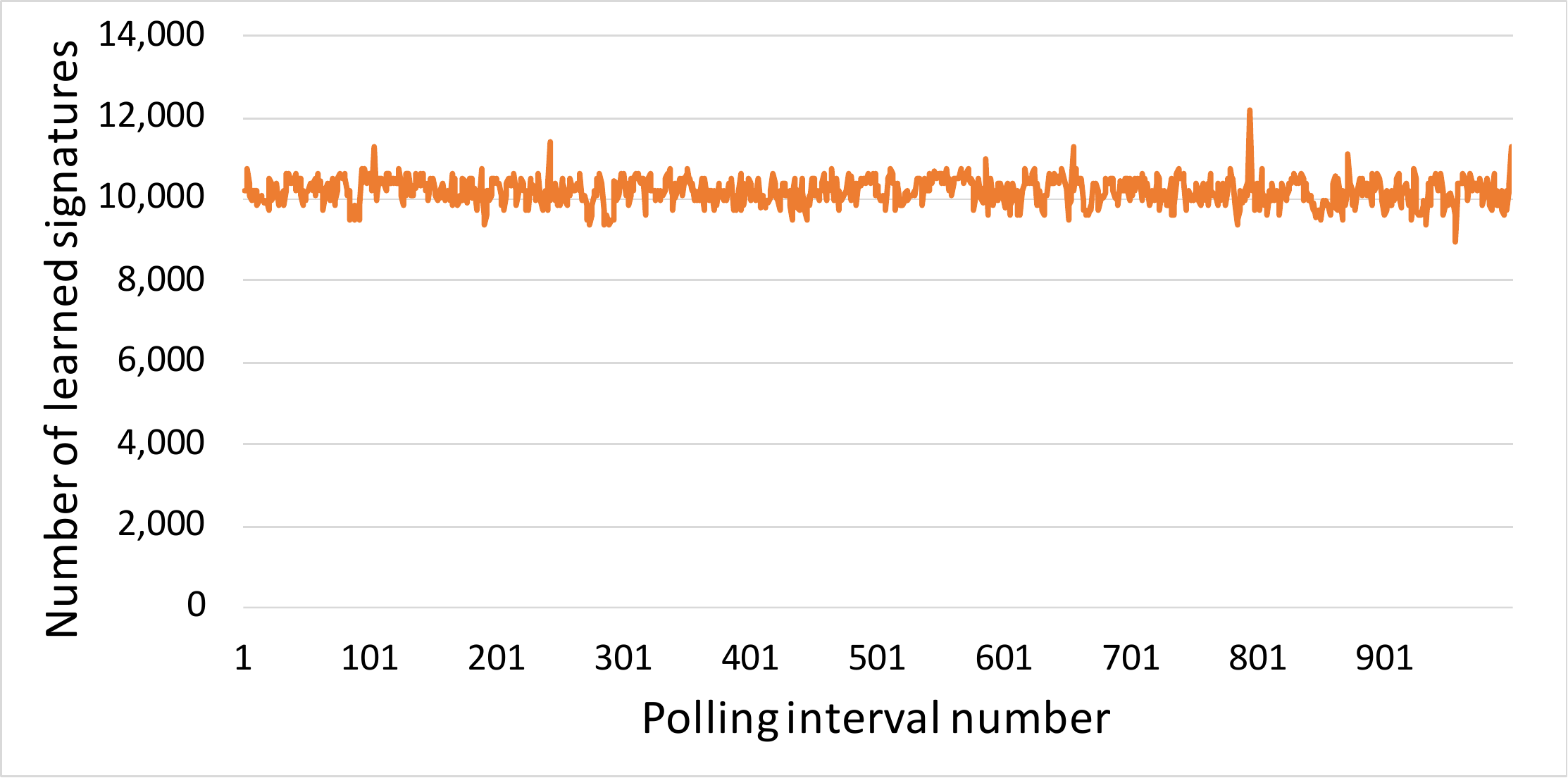}}
\caption{\centering The number of signatures learned during each 10$m$s polling interval when 1M SYNs are generated per second}
\label{fig:high_rate_learning}
\end{figure}

 Figure~\ref{fig:high_rate_learning} shows the number of signatures learned during each 10ms polling interval, for a 10-second experiment. The continuous high rate of 1M SYNs per second is generated using MoonGen~\cite{emmerich2015moongen}, a high speed packet generator built on DPDK~\cite{intel2014data}. As the graph shows, the FIB(SYN1) table is able to capture a burst of 10,000 SYNs per each 10$m$s polling interval, which is equivalent to 1 million new connections per second. The peaks and valleys in the graph  shows that the generated traffic does not have to be uniformly distributed, and that the prototype hardware can accommodate bursts of more than 12,000 SYNs per a 10ms polling interval.

\subsection{DIP pool updates}
\begin{figure}[t]
\scalebox{0.7}{\includegraphics[scale = 2,trim = 5cm 4cm 5mm 5cm, clip,width=0.8\textwidth]{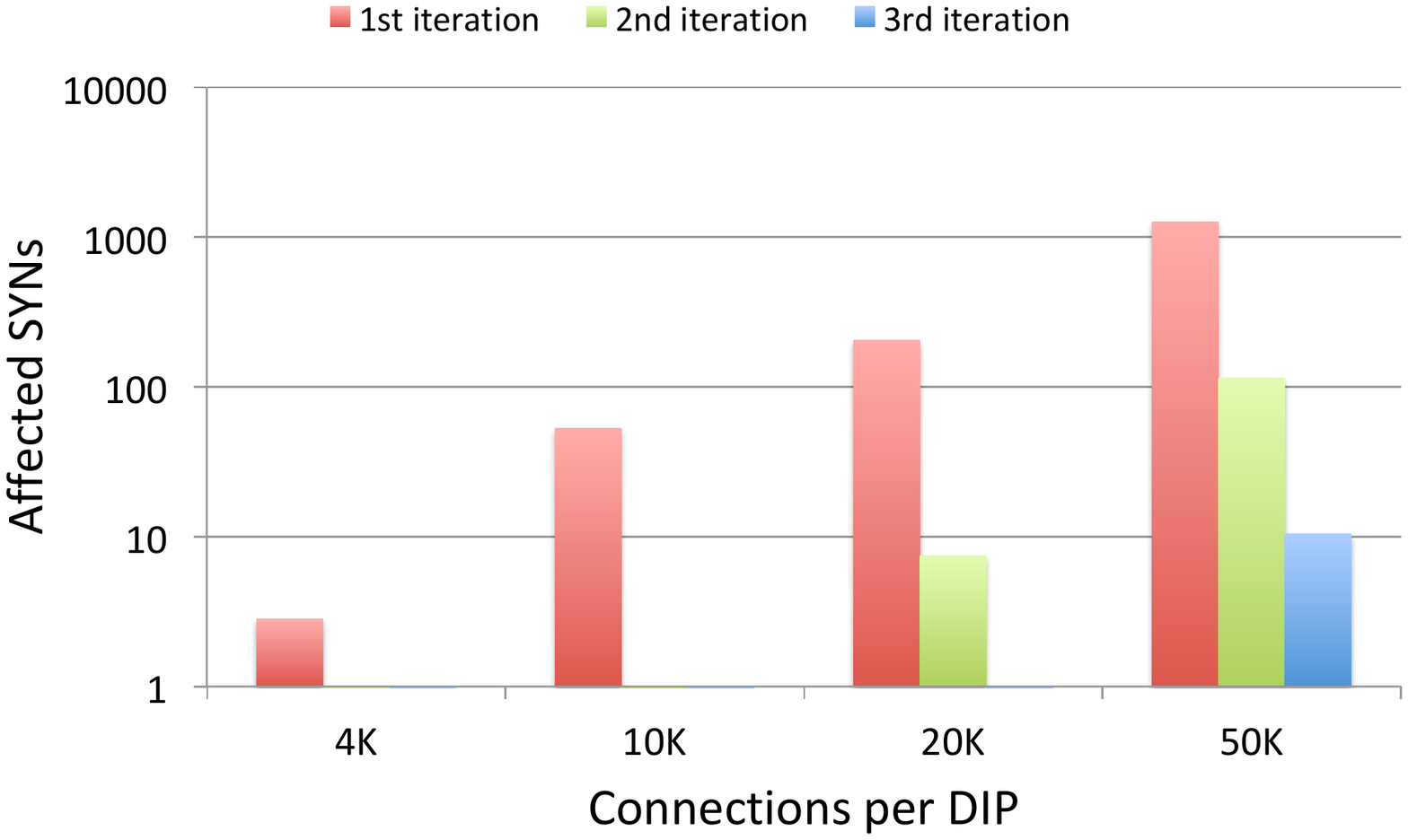}}
\caption{\centering New affected SYNs received during each iteration of Algorithm~\ref{alg:iterative migration} for the considered DIP }
\label{fig:connections_per_dip}
\end{figure}

Recall that in order to guarantee PCC, the MCT is used to maintain state for connections affected by DIP pool updates. The MCT entry processing is initiated by an administrative scale up or scale down notification, typically sent to the switch by an SDN controller. According to Algorithm~\ref{alg:iterative migration}, for each affected bin, the software reads the FIB(SYN1) table, updates the SCT and then writes the affected signatures into the MCT. Recall that during this time interval, additional new SYNs continue to arrive. They are stored by hardware FIB table, and are drained and processed in an iterative manner (see Algorithm~\ref{alg:iterative migration}).

\par If a single DIP is to handling up to 10K active connections, then Prism must manage around 10K DIPs, regardless of the number of VIPs to which these DIPs are assigned. In the POC implementation, we consider 1 VIP with 10K DIPs, as this makes the POC configuration and measurements easier. With 100K ECMP bins, there are 10 bins per DIP, each with 1,000 signatures. Thus, for each DIP change notification, Algorithm~\ref{alg:iterative migration} is executed 10 times, once for each bin. We measured  the total time for all the bins of the same DIP update  to be 450ms, which fits very well the mathematical analysis in Section~\ref{sec:dip pool update}. 
{\bf This implies that Prism can easily handle a pool change update once every 500ms, and 120 updates per minute.}

\par Figure \ref{fig:connections_per_dip} shows the number of SYNs received during each iteration of Algorithm~\ref{alg:iterative migration} for 4 different DIP configurations. When each DIP handles more active connections, the average time to program the MCT entries and handle the additional new connections grows accordingly. As shown in the graph, the second iteration is an order of magnitude smaller than the first, and we see that the algorithm converges for these profiles. Only for a DIP configuration of 50K, was a third iteration needed. 

\begin{figure}[t]
\scalebox{0.5}{\input{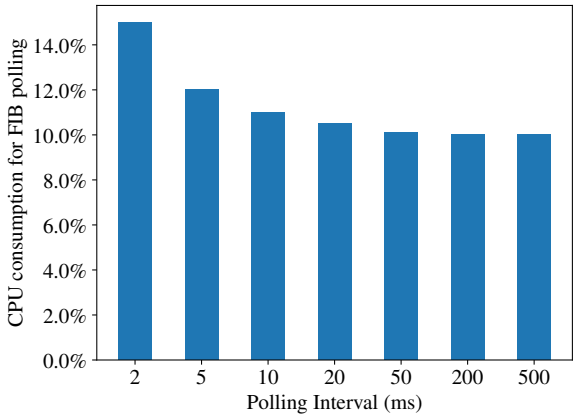}}
\caption{\centering The effect of the FIB polling interval on the switch's CPU consumption} 
\label{fig:fib_polling}
\end{figure}

\subsection{FIB polling}
Figure~\ref{fig:fib_polling} shows the effect of the FIB polling interval on the switch CPU consumption. 
The displayed CPU consumption is for the polling of all FIB tables. In this experiment, all FIB tables were polled together. Similar results are expected for polling the various table asynchronously. 

\par Consider the 10$m$s poll interval with 1M connections per second. During a 10ms interval, 10,000 connections start and end, on the average. Thus, 40,000 FIB entries with (SYN1, SYN2, FIN1, FIN2) have to be read. The rate of hardware read is 10M entries per second. Thus, 4ms are needed to complete the reading of 40,000 entries. There is also some hardware access
overhead of around 400$\mu$s, so the total time is 4.4$m$s. FIB reading is done by one processing core, which spends 4.4/10 = 44\% of its capacity. Since the total number of cores in Spectrum 2 is 4, Prism actually spends only 11\% of the total CPU on reading the FIBs. This is a small overhead, which leaves enough CPU resources for the other switch's tasks. 

\par From Figure~\ref{fig:fib_polling} we learn that the best trade-off between CPU consumption and polling interval is obtained for 10$m$s, because this is the smallest polling interval for which CPU consumption is close to the minimum of 10\%. 

\ifLongversion
\begin{figure}[t]
\scalebox{0.4}{\includegraphics{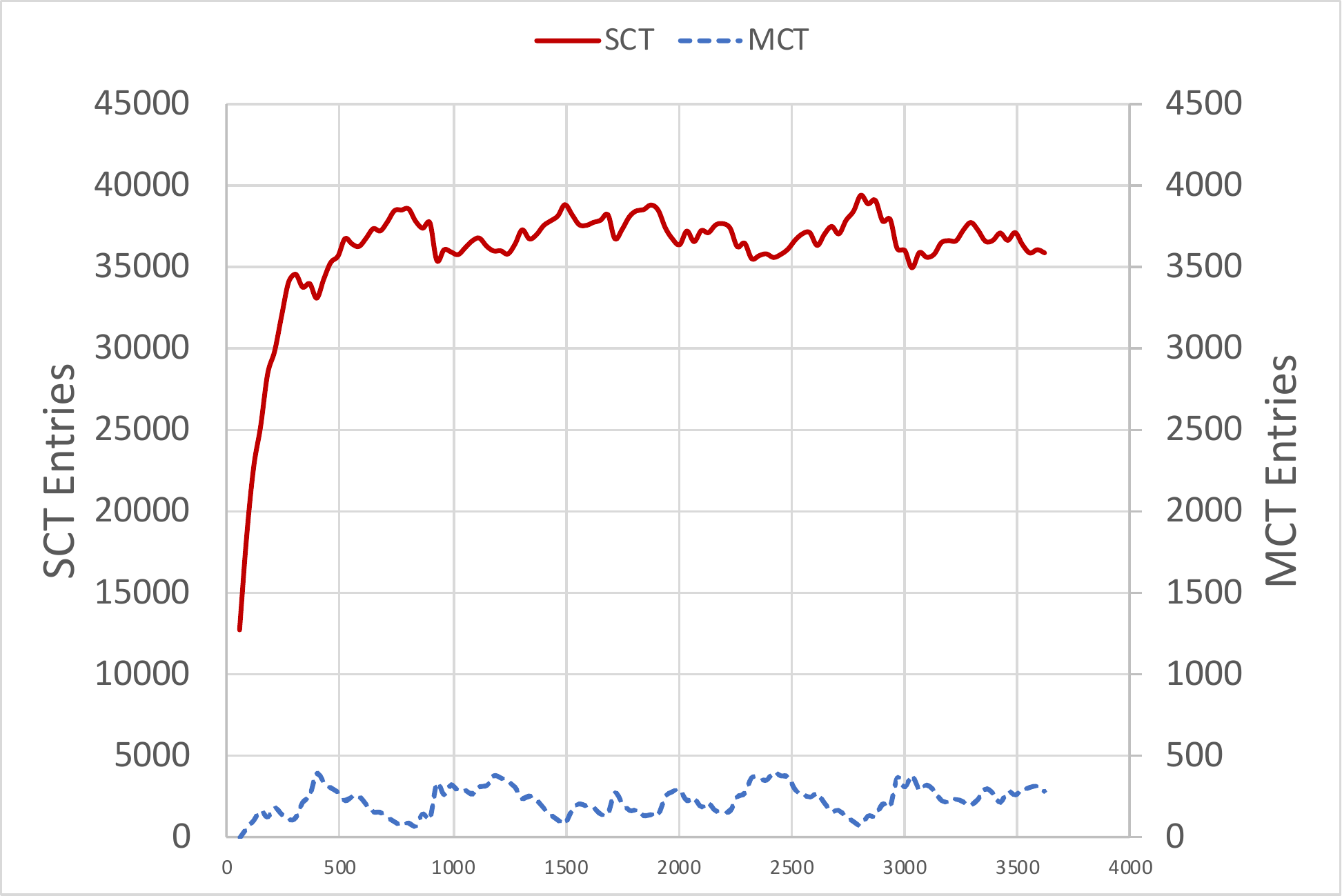}}
\caption{\centering MCT vs SCT size (the Y-axis for SCT is on the left and for MCT is on the right)}
\label{fig:mct_to_ct}
\end{figure}

\par We also measured the size of the MCT versus the size of the SCT under a constant average arrival rate of 1,000 connections per second to 100 DIPs. All connections are generated with a log normal distribution of connection lifetime, and an average lifetime of 40 seconds. In {\bf \color{blue} Figure~\ref{fig:mct_to_ct}}, the MCT to SCT  ratio stabilizes on 1:159, when 1\% of the system's DIPs are updated every minute, which is a very high pool update rate. {\bf \color{blue} (Weak PoC experiment, we reviewed this paragraph but we didn't include it in the submission (not even in the appendix). Mellanox should re-run the experiment with better parameters with the hardware learning on. The main issue is the very slow SYN arrival rate.)}
\fi

\ifLongversion



\par Table~\ref{tab:mct_measurments} shows the time taken to program the MCT entries and redistribute the ECMP bins of the migrated DIP, including the iteration to handle pending connections. The average time to create the MCT entries and poll for the incoming new connections was found to be 90$m$s.

\begin{table}[h!]
    \centering
    \begin{tabular}{ | m{16em} | m{5em} | }
    \hline
    Measured MCT entry create rate & 22K per sec \\ 
    \hline
    Initial time to process MCT + FIB & 530$m$s \\ 
    \hline
    New connections during update & 530K \\ 
    \hline
    New connections affected by DIP change & 53 \\ 
    \hline
    1st iteration MCT + FIB time & 10$m$s \\
    \hline
    New connections during 1st iteration & 9600 \\ 
    \hline
    2nd iteration MCT + FIB time & < 200$\mu$sec \\
    \hline
    New connections affected by DIP change & $\approx$1 \\ 
    \hline
    SYN to updated DIP sent to CPU & 0.4$m$s \\
    \hline
    Average number of SYNs per DIP change sent to CPU & 0.05 \\
    \hline
    \end{tabular}
    \caption{MCT measurements}
    \label{tab:mct_measurments}
\end{table}

\fi

\section{RELATED WORK}
\label{sec:related_work}
Several LB architectures have been proposed in recent years \cite{araujo2018balancing,eisenbud2016maglev,gandhi2015duet,miao2017silkroad,olteanu2018stateless,patel2013ananta,190437}. Duet \cite{gandhi2015duet} was one of the first papers to discuss cloud scale load balancing, and to leverage hardware features available in the switching ASICs. It uses the switch hardware to handle VIPs with high volume of traffic, leaving other VIPs to be handled by software. Duet utilizes the ECMP and tunneling hardware tables to maintain the VIP to DIP mappings. The number of entries in these tables is limited and cannot hold all the mappings. Thus, Duet saves additional complete mapping in the software, as Prism does. Duet handles VIP traffic using the hardware tables, and it forwards the traffic to the software when the destination VIP is being updated. In contrast, Prism maintains PCC while all packets are processed by hardware.

\par Rubik \cite{190437} aims to improve Duet by lowering the extra bandwidth incurred from traffic redirection to the software. It does so by using wise placement of DIPs in the datacenter. DIPs are placed closer to the source of the corresponding VIP traffic, and DIPs in the same VIP pool are placed close to the software LB handling traffic of the pool. In Prism, we deal with the more high level problem of improving the LB itself, rather than optimizing the topology of the datacenter.

\par SilkRoad \cite{miao2017silkroad} can balance the load imposed by ten million simultaneous connections using modern switching ASICs, for which we have already discussed at length in previous sections. \ifLongversion SilkRoad architecture includes a DIPPoolTable for mapping between (VIP, version) and DIP pools, a VIPTable for mapping between VIP and a version number, and a ConnTable for mapping between a connection signature and a version number. Incoming packets with VIP destination invoke the search process for the corresponding DIP. First, the ConnTable is checked according to the connection signature of the packet to get the version number. If the version number is available, the DIP is obtained from the DIPPoolTable. Otherwise, the newest version number can be obtained from the VIPTable. Using a 6-bit version field reduces the action data size, enabling SilkRoad to handle more connections in the hardware. 

\apr As in Prism, it is possible that two SilkRoad connections will have the same signature. Also as in Prism, to ensure PCC, SilkRoad uses TransitTable that remembers the set of connection being inserted into the ConnTable. Such connections should be mapped to an old DIP pool version during DIP pool update. TransitTable is implemented as a bloom filter, which does not need the CPU. SilkRoad presents promising concepts for Layer-4 load balancing, but it is still limited to handling only a few millions of ongoing connections, mainly due to the need to maintain a hardware state for each ongoing connection. \fi

\par Beamer \cite{olteanu2018stateless} leverages the per-connection state already maintained by the servers, without saving additional information about active connections. The main concept behind Beamer is that DIPs coordinate with each other and forward packets to each other when necessary. In Beamer, DIP pool updates cause packets to be routed to the wrong DIPs. To ensure PCC, DIPs that receive packets not belonging to them redirect them to the correct DIP. To implement this behavior, servers must be updated on the changes in the datacenter topology. This coordination is done using ZooKeeper \cite{hunt2010zookeeper}, which provides a scalable distributed storage and a notification service.

\par Faild \cite{araujo2018balancing} is a distributed load balancer, which also runs on commodity hardware. It achieves graceful failover without relying on network state. Like Beamer, Faild utilizes existing DIP's information to determine if a received packet should be redirected to another DIP.

\par Maglev \cite{eisenbud2016maglev} and Ananta \cite{patel2013ananta} are software LBs that run on commodity servers. Maglev is Google’s network LB, and Ananta is deployed in the Azure public cloud. Both Maglev and Ananta are not designed to benefit from modern ASICs capabilities, and the fact that they are software LBs results in high consumption of compute resources, additional connection latency and limited total throughput.

\vspace{-0.1cm}
\section{Conclusions}
\label{sec:conclusion}
This paper proposed a highly scalable LB, called Prism, implemented using a programmable switch ASIC. Prism is the first reported LB that can handle millions of connections per second and hundreds of millions connections in total, while ensuring PCC. Prism processes all the packets in hardware, even during server pool updates, and does not need to maintain a hardware state per active connection. We implemented a prototype of the proposed architecture, and proved that it can indeed accommodate 100M connections, with an arrival rate of 1M connections per second.

\clearpage
\bibliographystyle{plain}
\bibliography{references}

\clearpage
\appendix
\appendixpage
\section{Choosing Bins for Migration}
\label{sec:choosing bins}
Section~\ref{sec:dip pool update} mentioned three different events related to DIP pool change for which signature migration from the SCT to the MCT is needed: (a) graceful take down of a DIP; (b) adding a DIP; (c) changing the load between DIPs. This section addresses the question which bins should be chosen for migration.

\par Consider the first event (a), and suppose that there are 64 bins that contain the identity of DIP1 and 64 bins that contain the identity of DIP2. Suppose also that one of these DIPs has to be removed from the pool because connection arrival rate becomes slower. Thus, Prism can choose either to populate the bins that currently contain the identity of DIP2 with the identity of DIP1 and to take down DIP2, or vice versa. Similar decisions have to be made with respect to (b) and (c).

\par The decision for which bins to choose should not be made randomly. The reason is that each bin chosen for replacing its content is used by a certain number of active connections, and the signatures of all these connections must be copied into the MCT using Algorithm~\ref{alg:iterative migration}. Since the MCT is a scarce resource, Prism seeks to minimize the number of signatures it contains. One way to do this is to choose the bins that are used by the minimum number of signatures. The information for making such a decision exists in the SCT.   

\par However, not every migrated signature has the same impact on the MCT. A signature that stays in the MCT longer than another has a higher migration cost. Thus, when choosing a bin for migration, we need to take into account the duration each signature is expected to stay in the MCT. This is especially true for long lived connections, such as those that are established with a video server or file transfer. For this reason, Prism distinguishes between "regular VIPs", whose connections are relatively short, and "special VIPs", whose connections are long. The classification of VIPs is conducted using statistics obtained by Prism from the SCT.

\par For regular VIPs, bin selection is performed while minimizing the number of affected signatures. For the special VIPs, Prism seeks the bins whose signatures will minimize the total expected load imposed on the MCT. For each $bin_i$, the expected load of its signatures is defined as:
$$W(bin_i) = {\sum_{SIG_j \in SIG(bin_i)}T - time(SIG_j) } ,$$
where $T$ is the expected time duration of a connection for the considered VIP, and $time(SIG_j)$ is the time elapsed since $SIG_j$ has been received; thus, $T - time(SIG_j)$ is the expected time the signature will stay in the MCT. If Prism needs to choose $n$ bins for migration, it chooses those for which $W(bin)$ is the minimum. Algorithm~\ref{alg:choose bins} summarizes this idea.

\SetKwInput{KwInput}{Input}                
\SetKwInput{KwOutput}{Output}              
\SetKwInput{KwStart}{Start}              
\begin{algorithm}
\SetAlgoLined
\KwInput{\\
$n$: the number of bins to be chosen for migration;\\
$bin_1 \dots bin_k$, where k > n: the candidate bins;\\
}
\KwStart{}
  $W_{bins} \gets \{\}$;\\
 \For{$i\gets0;\ i \leq k;\ i\gets i+1$}{
        $W_{bins} \gets W_{bins} \cup \{(bin_i,W(bin_i))\}$;\\
 }
 sort $W_{bins}$ by $W(bin)$, and return the minimum n bins;\\
 \caption{choosing bins for migration}
 \label{alg:choose bins}
\end{algorithm}

\begin{figure*}[ht!]
\centering
\begin{minipage}[t]{.4\textwidth}
\scalebox{0.45}{\input{simulation_figures/frequent_updates_bins_algorithm_small.pgf}}
\caption*{\small \textmd{(a) average connection length = 50s}}
\end{minipage}\qquad\qquad
\begin{minipage}[t]{.4\textwidth}
\scalebox{0.45}{\input{simulation_figures/frequent_updates_bins_algorithm_big.pgf}}
\caption*{\small \textmd{(b) average connection length = 500s}}
\end{minipage}
\caption{\centering The percentage of connections of the considered VIP which are also stored in the MCT throughout system running time, as function of the method for choosing bins}
\label{fig:choosing bins}
\end{figure*}

\par To examine the impact of Algorithm~\ref{alg:choose bins} on the system performance, we simulated a VIP that receives 10,000 connections per second. The pool for this VIP has 1,000 DIPs, each residing in a bin. The expected connection lifetime for the considered VIP is 50 (Figure~\ref{fig:choosing bins}(a)) or 500 (Figure~\ref{fig:choosing bins}(b)) seconds. The experiment consists of removing 1/4 of the DIPs due to reduced incoming connection rate for this VIP. Bins are taken down one by one, and their contents are replaced with new DIPs from outside the pool.

\par Figure~\ref{fig:choosing bins} shows the simulation results for three cases: when Algorithm~\ref{alg:choose bins} is used for choosing the affected bins, when a non-weighted version of Algorithm ~\ref{alg:choose bins} is used (i.e., the least populated bins are chosen), and when the bins are chosen randomly. The total load imposed on the MCT by the considered pool update is the integral of each curve. In Figure~\ref{fig:choosing bins}(a), the load on the MCT when Algorithm 2 and non-weighted are used is smaller by 15\% and 13.2\%, respectively, compared to random. In Figure~\ref{fig:choosing bins}(b), the improvement is 35\% and 29.1\% respectively.  

\section{P4 Implementation}
\label{app:p4_implementation}

From Figure \ref{fig:hybrid_pipeline}, the main switch pipeline stages consist of the following. At the flex1 stage we place a Load Balancer Control Table (LBCT), which is responsible for matching TCP packets with the SYN/FIN/RST flags enabled:
\begin{lstlisting}
table tcp_trap {
    key = {
        headers.ip.ipv4.protocol : exact;
        headers.ip.ipv4.src_addr : ternary;
        headers.ip.ipv4.dst_addr : exact;
        headers.tcp.src_port     : ternary;
        headers.tcp.dst_port     : exact;
        headers.tcp.flags        : ternary;
    }
    actions = {
        no_action;
        conn_signature;   /*parameter table_id */
        learn_signature;
        cpu_trap;
    }
    default_action = conn_signature();
}
\end{lstlisting}
 
\par If Control Match Action Table is matched, the action is to create the connection signature and pass the packet to the FIB table to be learned. Other actions include special handling for exception packets, where we want the packet trapped and sent to the CPU.

The FIB table is based on a fixed hardware unit for MAC learning. It is reused as the connection learning table (bridge stage in Figure \ref{fig:hybrid_pipeline}), and does not need to be modeled in P4. FIB table stores the new connection signatures and interacts with the software polling thread.
 
\begin{figure*}[p]
\vspace{-7in}
\centering
\begin{minipage}[t]{.4\textwidth}
\hspace{-2cm}
\scalebox{2}{
\includegraphics[trim = 3cm 5cm 5mm 6cm, clip,width=0.8\textwidth]{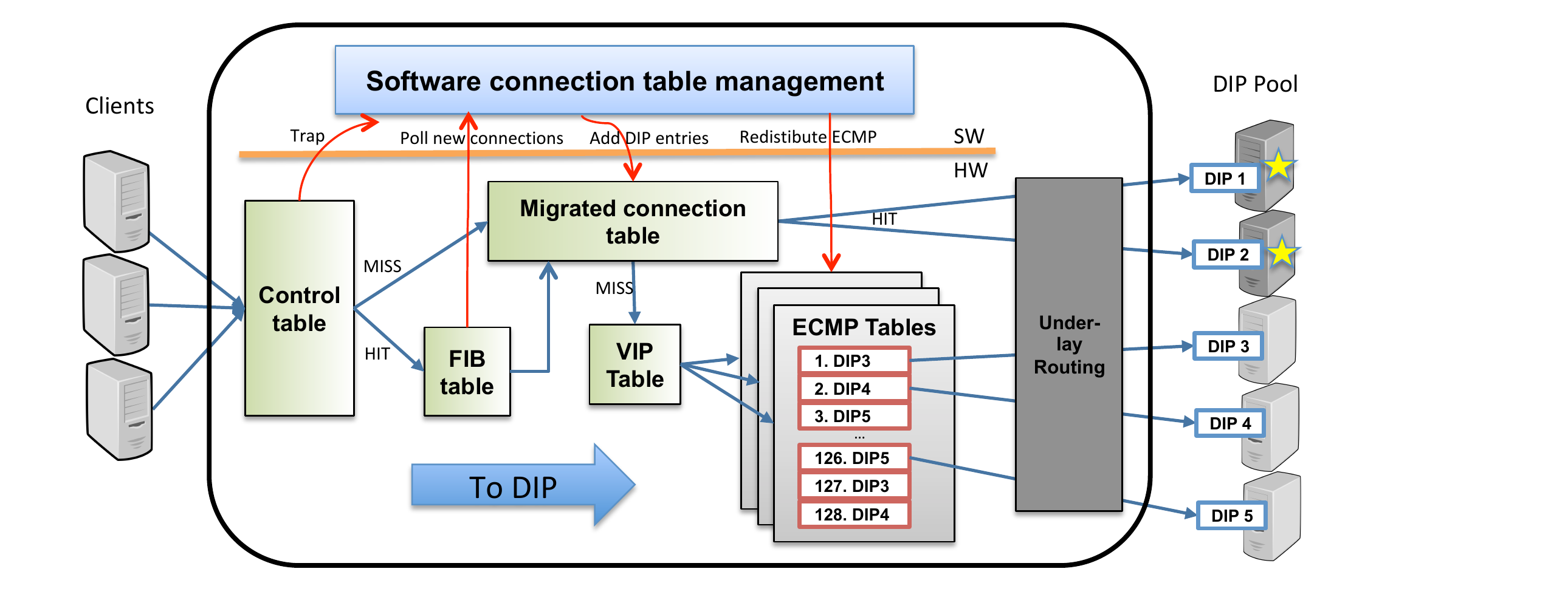}}
\caption*{\small \textmd{(a) Client to Server path}}
\end{minipage}\qquad\qquad
\begin{minipage}[t]{.4\textwidth}
\scalebox{2}{
\includegraphics[trim = 6cm 6cm 5mm 6cm, clip,width=0.8\textwidth]{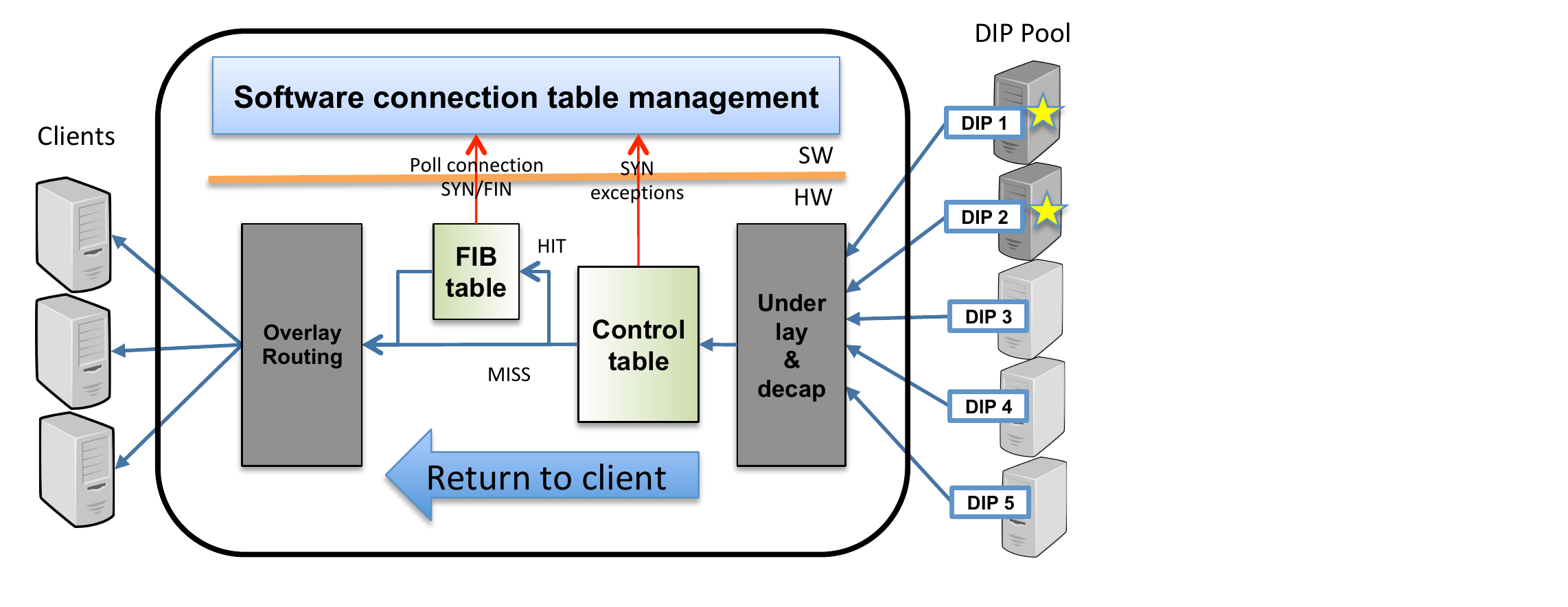}}
\vspace{-0.5cm}
\caption*{\small \textmd{(b) Return path}}
\end{minipage}
\caption{\centering P4 tables and DIP packet path }
\label{fig:flex_tables}
\end{figure*}

On an LBCT miss, the default action is to proceed to the MCT table (flex2 stage in Figure \ref{fig:hybrid_pipeline}). A match/hit on an MCT table entry will direct the packet to the routing block, where it proceeds directly to the reassigned DIP. For example, see DIP1 and DIP2 in Figure~\ref{fig:flex_tables}.
\begin{lstlisting}
table migrated_connection {
    key = {
        metadata.conn_signature  : exact;
    }
    actions = {
        no_action; to_nexthop;
    }
    default_action = no_action();
    size = 4096;  /* minimum size */
}
\end{lstlisting}
 On an MCT miss, the VIP table is then applied to select which ECMP table resource should be used for the actual load balancing.
\begin{lstlisting}
// calculate signature 
action set_next_ecmp()
{
    hash(
        metadata.register,  /* output result */
        HashAlgorithm.crc,  /* input hash algo */
        1w0,                /* input base */
        {
            metadata.conn_signature,
        },              /* input data */
        64w0xFFFFFFFF   /* input 64 bits max */
    );
}
action_selector(HashAlgorithm.crc, 4096 /*size*/, 
                64 /*width*/) as;
table ecmp {
    key = {
        metadata.register : selector;
    }
    actions = {
        set_next_ecmp;  no_action;
    }
    size = 4096;       /* minimum size */
    implementation = as;
}
\end{lstlisting}

 The hash function is configurable and chosen to be based on the TCP/IP 5-tuple. We size each table by assigning roughly 10 bins to each DIP. For 10K DIPs, this means 100K ECMP bins. After the ECMP block is evaluated, the packet proceeds to the legacy underlay routing. This entire custom P4 flow is shown in Figure~\ref{fig:flex_tables}(a) for packets going from the client IP towards the DIPs. All packets returning from the server towards the clients are also forwarded by the switch, but in this direction do not need to go through the ECMP tables (see Figure~\ref{fig:flex_tables}(b)). In Figure~\ref{fig:flex_tables}(b) there is a legacy underlay/decap block, followed by LBCT and FIB learning, so the software can maintain the connection state from the SYN/FIN packets originated by the DIPs.



\ifLongversion

\fi
\end{document}